\documentclass[manuscript,screen]{acmart}

\AtBeginDocument{%
  }

\setcopyright{acmlicensed}
\copyrightyear{2026}
\acmYear{2026}
\acmDOI{XXXXXXX.XXXXXXX}
\acmISBN{978-1-4503-XXXX-X/2018/06}

\begin{document}

\title{A Tertiary Review of Large Language Model-Based Code Generating Tasks: Trends, Challenges, and Future Directions}

\author{Muslim Chochlov}
\email{muslim.chochlov@ul.ie}
\affiliation{%
  \institution{University of Limerick}
  \city{Limerick}
  \country{Ireland}
}

\author{Michael English}
\affiliation{%
  \institution{University of Limerick}
  \city{Limerick}
  \country{Ireland}
}
\email{michael.english@ul.ie}

\author{Jim Buckley}
\affiliation{%
  \institution{University of Limerick}
  \city{Limerick}
  \country{Ireland}
}
\email{jim.buckley@ul.ie}






\renewcommand{\shortauthors}{Trovato et al.}

\begin{abstract}

\textbf{Context.} Large language models (LLMs) are increasingly applied to code-generating tasks (CGTs) in software engineering. While reported results are promising, the broader effects of such application and their integration into real-world development remain insufficiently understood with existing tertiary studies provide little in this area.

\textbf{Objective.} This tertiary study consolidates secondary evidence on LLM-based CGTs, synthesizing the publication landscape, effects, scenarios, integration challenges, and future research directions.

\textbf{Method.} Following systematic review guidelines, we searched in related digital libraries, complemented by backward-and-forward snowballing and screening step. Study quality was assessed and extraction reliability was audited with inter-rater agreement statistics. Evidence was synthesized using SWEBOK knowledge areas and the HELM framework.

\textbf{Results.} We identify 30 secondary studies published between 2017-2025, with rapid growth since 2023. Accuracy seems strong on benchmarks but weakly supported for real-world generalization; robustness is fragile across tasks and configurations; efficiency constraints are pervasive; toxicity and bias are under-reported. Dominant challenges concern economic feasibility, evaluation validity, and socio-technical integration. Future directions suggest domain-aware model improvement and the need for holistic, standardized evaluation.

\textbf{Conclusion.} LLM-based CGTs represent a fast-maturing yet unevenly evaluated research area, highlighting the need for domain-aware model improvements and holistic, standardized evaluation, addressing efficiency and associated costs.
\end{abstract}

\ccsdesc[500]{Software and its engineering~Maintaining software}
\ccsdesc[500]{Computing methodologies~Artificial intelligence}
\keywords{systematic literature review, large language models, software engineering, code generating tasks}


\maketitle

\section{Introduction}
Recently, software engineering (SE) has become increasingly supported by contemporary large language models (LLMs) that can, among other tasks, generate code artifacts from natural language, code context, specifications, or other inputs \citep{hou2024large}. This seems to bring efficiencies: for example, Peng et al. report that programmers employing Github Copilot completed the task 55.8\% faster than the control group \cite{peng2023impact}. Other evidence, coming from trials at Microsoft, Accenture, and a Fortune 100 company suggests a 26.08\% increase in completed coding tasks when AI is equipped \cite{cui2025effects}. Likewise, in code competitions requiring code generation, AlphaCode has achieved average rankings in the top 54.3\% of evaluations \cite{li2022competition}. These results indicate that LLM-based CGTs are not merely incremental improvements over prior techniques, but represent a shift toward broadly applicable, generative AI-driven development support. Yet, despite these promising results the current applicability of LLMs towards code generating tasks (CGTs) (i.e., software engineering tasks whose primary outcome is automatically produced executable source code, such as code generation, test generation, or program repair) that go beyond code snippet generation/code competitions is questionable \cite{hou2024large}: particularly, the effects and challenges of applying such LLMs for real-world code generating scenarios are still under-reported. 

The existing related secondary studies and tertiary reviews are either too broad (e.g. looking at the intersection of machine learning (ML) and SE \cite{kotti2023machine}), are focused only on a subset of code generating tasks (e.g. looking at AI in software testing \cite{amalfitano2023artificial}), or are narrowly focused on cross-task aspects of AI4SE like prompt engineering \cite{tony2024prompting}. This makes it difficult to isolate the state of evidence specifically for LLM-based CGTs.


To address this gap we synthesize evidence from existing secondary studies in our tertiary review with the goal of discovering the existing landscape, effects of applying LLM-based CGTs mapped to HELM (Holistic Evaluation of Language Model: framework designed for holistic evaluation of LLMs for natural language tasks\cite{liang2022holistic}) measures and scenarios, existing integration challenges, and possible future directions. We follow Kitchenham’s SLR guidance and SEGRESS reporting \cite{kitchenham2007guidelines, kitchenham2022segress}. The protocol (search strings, screening rules, extraction forms, and analysis scripts) and the curated dataset are publicly released for transparency and reuse \cite{zenodo}. Our search spans multiple recommended digital libraries \cite{gusenbauer2020academic} and Google Scholar, combines database querying with structured snowballing \cite{wohlin2014guidelines}, and incorporates a validated semi-automated screening step using a state-of-the-art LLM in line with emerging evidence-synthesis best-practice \cite{wallace2010semi,o2015using}. Reliability safeguards include multi-rater audits (Fleiss’ $\kappa$ and Cohen’s $\kappa$) and an extractor–checker strategy for qualitative evidence extraction and synthesis with Wilson 95\% confidence intervals for the resulting proportions, following Kitchenham’s recommendations for secondary studies in SE.

We structure the paper around the following research questions (RQs):
\begin{enumerate}  
\item \textbf{RQ1}: What is the landscape of secondary studies that report (at least in part) on LLM-based CGTs?
Sub-questions:
\begin{enumerate}
    \item What is the distribution of secondary studies by year and by study type (e.g. mapping studies/systematic reviews)?
    \item How many primary studies are reviewed in these secondary studies?
    \item In what venues are they published?
    \item What is AI and SE domain scopes of these studies?
\end{enumerate}
\item \textbf{RQ2}: What HELM measures of LLM-based CGTs are reported in secondary studies? 
\item \textbf{RQ3}: What HELM scenarios of LLM-based CGTs are reported in secondary studies?
\item \textbf{RQ4}: What challenges in applying or integrating LLM-based CGTs into software engineering workflows are identified in secondary studies?
\item \textbf{RQ5}: What future directions for LLM-based CGTs are reported on in existing secondary studies?
\end{enumerate}


This paper makes the following contributions:
\begin{itemize}
\item \textbf{Tertiary synthesis of LLM-based CGTs.} Using a Kitchenham/SEGRESS-compliant protocol, we identify 30 secondary studies (2017–2025), quantify the publication landscape (year, study type, venue), map their AI and SE scopes, map the observable effects to the adopted HELM \cite{liang2022holistic} framework, and map their challenges and future directions against our derived (bottom-up) taxonomies grounded in existing frameworks \cite{SWEBOK2024, liang2022holistic, kitchenham2007guidelines}.
\item \textbf{Novel application of LLM towards literature review.} We describe and show how LLM (GPT4o) can be used reliably to improve efficiency of snowballing (foe e.g. recursively reviewing large lists of references and citations) as part of a literature review process, potentially saving human-hours.
\item \textbf{Novel adoption of the HELM framework.} We describe and show how the HELM framework (originally designed for natural language based LLM holistic evaluation) can be adapted to SE contexts by retaining its multi-measure design and redefining scenarios to reflect SE-specific dimensions (task, programming language, application type/domain), consistent with SWEBOK v4 and ISO/IEC~25010 \citep{liang2022holistic, SWEBOK2024, iso25010}.
\end{itemize}


This paper is structured as follows: Section~\ref{sec:def-cgts} provides CGT/LLM definitions, describes the employed HELM framework and discusses related work. Section~\ref{sec:methodology} details the methodology (RQs, search, selection, quality/bias assessment, data extraction and synthesis). Section~\ref{sec:results} reports results and answers for RQ1–RQ5, including the landscape, HELM-aligned effects, scenarios, challenges, and future directions with Section~\ref{sec:discussion} providing discussion of these results. Section~\ref{sec:threats} discusses threats to validity. Section~\ref{sec:conclusions} provides a conclusion to this work.

\section{Background and Related Work}
\subsection{Code Generating Tasks}
\label{sec:def-cgts}

Existing research usually treats code-producing tasks either in isolation (e.g., focusing on code generation, test generation, automated program repair \cite{jiang2024survey, zubair2025use, wang2024software}) or in a broader context of software engineering \cite{hou2024large}, which leaves the field without a shared definition for/vision of code-generation  tasks. Therefore, here we adopt a unifying umbrella definition for scoping these tasks in this work:

\emph{Code-generating task (CGT)} is a software engineering task whose primary output is an automatically generated code artifact written in a programming language that is intended to be compiled or interpreted and executed (such output can include tests as code and repair patches). The input for CGTs varies and can include natural language, existing code, documentation, images, and other types of input \cite{hou2024large, ahmed2023source}. (Tasks that solely produce non-executable text (e.g., inline code comments, docstrings, or documentation) are not considered a CGT.) Scale of code produced by CGTs varies as well and can be as small as line/token level (e.g. in code completion tasks) or as large as package/system level (e.g. in program synthesis task).

To the best of our knowledge, there is no finite set of CGTs defined in the literature, although several studies provide attempts at compiling comprehensive lists of software engineering tasks\cite{hou2024large, zheng2023survey}. Based on these lists, augmented by existing LLM-based code-generating literature, a set of CGTs tasks can be constructed. Below we give the commonly used names for these CGTs, their input/output (input $\rightarrow$ output), and short descriptions as they appear in the literature.
\begin{itemize}
  \item \textbf{Code generation (natural language/examples/formal specifications $\rightarrow$ code).} Produces new code, ideally consistent with user intent \cite{chen2021evaluating}.
  \item \textbf{Program synthesis (natural language/examples/formal specifications $\rightarrow$ code).} Similar to code generation, but at larger scale \cite{gulwani2017program}.
  \item \textbf{Code completion (code context $\rightarrow$ next code lines/blocks).} Edit-time, multi-token, but small-scale code generation, conditioned on local code context \cite{svyatkovskiy2020intellicode, chen2021evaluating}.
  \item \textbf{Patch generation as part of program repair / vulnerability repair (code $\rightarrow$ repaired code).} Generates code patches that are intended to pass defined tests or fix reported defects / vulnerabilities \cite{monperrus2014critical, liu2021critical}.
  \item \textbf{Test generation (natural language/specifications/code $\rightarrow$ tests as code).} Intends to produce unit / integration / other forms of tests as code \cite{fraser2011evosuite,fraser2012whole, schafer2023empirical}.
  \item \textbf{Code translation (code $\rightarrow$ code).} Migrates code across languages while preserving functionality \cite{lachaux2020unsupervised}.
  \item \textbf{Refactoring (code $\rightarrow$ functionality-preserving code).} Applies functionality-preserving transformations (e.g. rename / re-structure / move / de-duplicate / optimize) to code \cite{baqais2020automatic}.
\end{itemize}

\subsection{Large Language Models}
Contemporary large language models (LLMs) are commonly defined in terms of their architecture, training state, and their size. In terms of the architecture, these models are commonly built using the “Transformer’’ \emph{architecture}, a deep multi-layered neural network, introduced by Vaswani et al \cite{vaswani2017attention, hou2024large}. Particularly, this architecture facilitattes training parallelism, allowing models to train faster and to scale with more data and parameters. Also, its self-attention mechanism allows it to better capture relationships between (distant) input tokens and to better understand their context. Common subtypes of this architecture are encoder, decoder and encoder-decoder, where the latter two are predominantly used for generative tasks \cite{wang2024software}.

Transformer-based models commonly begin with a self-supervised pre-training stage: the model learns general knowledge from large unlabeled corpora of natural or programming language(-s). This is achieved by solving objectives, most commonly \emph{next-token prediction} with a decoder-only transformer (e.g. GPT), \emph{masked-token prediction} with an encoder (e.g. BERT), or \emph{denoising} in an encoder-decoder (T5/BART) \cite{devlin2019bert,raffel2020exploring,lewis2019bart,brown2020language}. Afterward, models can be adapted by fine-tuning, including alignment methods such as instruction tuning \cite{peng2023instruction} and reinforcement learning from human feedback (RLHF) \cite{ouyang2022training}.

The size of these models is usually defined by the number of their parameters. At the moment, there seems to be no formal consensus on a parameter threshold that would separate 'large' language models from language models. Existing research notes that usage is contextual, with many works informally treating models \emph{larger than $\sim$10B parameters} as large, while others emphasize ``tens to hundreds of billions'' \citep{zhao2023survey,minaee2024large}. However, some other research treats models with as little as 100M parameters as large \cite{wang2024software, chochlov2022using}: in the real-world language models span orders of magnitude: BERT (125M),  LLaMA (7–70B), GPT-3 (175B), and PaLM (540B) \cite{brown2020language,chowdhery2023palm,touvron2023llama,devlin2019bert}. 

Therefore, for all practical reasons of this paper, we can define LLMs as Transformer-based (with decoder or encoder-decoder components for generative tasks) pre-trained language models, trained on broad corpora of code with self-supervised objectives and having at least 100M parameters. Subsequently, these models can be adapted by fine-tuning (e.g., instruction tuning/RLHF).

\subsection{Adapting the HELM Framework for Software Engineering Contexts}
\label{subsec:helm_framework}
To the best of our knowledge, no existing taxonomy has been tailored specifically for LLM-based CGT evaluation. Therefore, here we adopt and adapt the Holistic Evaluation of Language Models (HELM) framework \cite{liang2022holistic} because it offers a holistic, multi-dimensional evaluation methodology that is increasingly recognized in the AI community towards LLM evaluation, originally targeted at natural language tasks. Our adaptation preserves HELM’s design rationale while aligning its constructs with the factors unique to CGTs. 

The framework~\cite{liang2022holistic} supports evaluation of LLMs through two main components: \emph{scenarios} and \emph{measures}.

\paragraph{Measures.} 
HELM defines a set of 7 evaluation measures capturing aspects of model's performance. Table~\ref{tab:helm-measures-adapted} summarizes these measures and shows their CGT-specific interpretation. In applying HELM to code-generating tasks, we interpret its dimensions at the level of CGT usage in software engineering workflows, rather than restricting them exclusively to model-internal properties or final software artifacts.

\begin{table}[h]
\centering
\caption{Mapping HELM measures to CGT contexts.}
\resizebox{\textwidth}{!}{\begin{tabular}{p{3cm}p{5.5cm}p{5.5cm}}
\toprule
\textbf{HELM Measure} & \textbf{Summary of Original Definition} & \textbf{Interpretation for CGT Context} \\
\midrule
Accuracy & Task correctness vs.\ ground truth &
Functional correctness of generated code (e.g., tests passed, specification compliance) \\
Calibration & Confidence vs.\ correctness alignment &
Alignment between model confidence and actual correctness of code suggestions \\
Robustness & Stability under perturbation &
Stability of model behavior under incomplete, ambiguous, or adversarial prompts during code generation \\
Efficiency & Resource usage and latency &
Computational efficiency of code generation (latency, cost), and—when deployed—its impact on developer workflow (e.g., IDE responsiveness) \\
Toxicity & Offensive or harmful natural language &
Generation of insecure code patterns, vulnerabilities, or license-violating fragments \\
Fairness & Human group related performance disparities &
Disparities across developer groups or contexts (e.g., experience level, programming background) \\
Bias & Undesirable associations/stereotypes &
Systematic preference for particular coding styles, APIs, frameworks, or programming languages \\
\bottomrule
\end{tabular}
}
\label{tab:helm-measures-adapted}
\end{table}

As a result, different HELM dimensions naturally map to different evaluation targets: some primarily reflect properties of the generated code (e.g. Accuracy), others properties of the underlying model behavior (e.g. Robustness), and others the interaction between the model and the development process (e.g. Efficiency).

\paragraph{Scenarios.} 
HELM defines a \emph{scenario} as a triple (task, domain, and language), further dividing domain into (what (e.g. text), who (e.g. author), when (e.g. some point in time)). While we retain the most of HELM’s scenario's structure, the original ``what/who/when'' decomposition of \emph{domain} is tailored to natural language contexts and does not directly extend to code. Likewise, in SE, other factors such as the \emph{application domain} (e.g., embedded systems, enterprise software, scientific computing) and the \emph{application type/system type} (e.g., mobile application, web service, safety-critical system) are more descriptive than 'domain'. Our adaptation therefore defines scenarios as follows:
\begin{itemize}
    \item \textbf{Task:} Retained, but as CGTs such as code generation, program repair, vulnerability repair, and test generation, as per the definition above.
    \item \textbf{Language:} Retained, referring to the programming language of generated artifacts (e.g., Python, Java, C++).
    \item \textbf{Domain:} Reinterpreted as a couple of \emph{application domain} and \emph{application type/system type}, consistent with SE standards such as SWEBOK v4 \cite{SWEBOK2024} and ISO/IEC~25010 \cite{iso25010}.
\end{itemize}

With these adaptations, we preserve the methodological structure of the original framework while extending its applicability to CGTs.

\subsection{Related Work}
\label{sec:rw-llm-cgts}
Tertiary systematic literature reviews (SLRs) on the intersection of artificial intelligence (AI) and software engineering (SE) research areas are few. Here, we present two such existing tertiary reviews and highlight their differences with our work in Table~\ref{tab:venue_summary}. Amalfitano et~al. synthesize secondary studies at the intersection of AI and software testing (ST) with studies collected starting from 1995 and going to 2022, drawing on established taxonomies from both areas (AI/ST) to map 20 reviews and to characterize popular applications and research gaps \cite{amalfitano2023artificial}. Independently, Kotti et~al. aggregate 83 secondary studies on machine learning (ML) for SE, starting from 2009 and going to 2022, reporting that quality and testing dominate in terms of the SE area and outlining challenges such as stronger empirical validation, better data pipelines, and more industrial studies \cite{kotti2023machine}. 

Our work significantly differs from these studies, especially in AI scope, SE scope, and the years for which publications were analyzed. On the AI scope, we narrow from AI/ML in general to LLMs. On the SE scope, we move from the entirety of SE (or the single Knowledge Area of software testing, as defined in SWEBOK v4 \cite{SWEBOK2024}) to focus on a set of CGTs (somewhat orthogonal to a Knowledge Area) whose primary outputs are intended as compilable/executable/ interpretable code artifacts (code generation, completion, program repair, test generation, translation, refactoring - see Section~\ref{sec:def-cgts}). Finally, in terms of years for which studies were collected, we look at the period (2017 - 2025) when contemporary LLM-based approaches started to appear en-masse as opposed to older approaches reviewed by Amalfitano and Kotti (see Table~\ref{tab:rw-positioning}). This tighter focus enables us to concentrate on contemporary LLM-based CGTs and to synthesize data for these approaches that prior broad studies could not emphasize.

\begin{table}[t]
\centering
\caption{Positioning relative to closely related tertiary studies.}
\label{tab:rw-positioning}
\resizebox{0.5\textwidth}{!}{\begin{tabular}{@{}lllr@{}}
\toprule
\textbf{Author} & \textbf{AI scope} & \textbf{SE scope} & \textbf{Years}\\
\midrule
Amalfitano et al.\ \cite{amalfitano2023artificial} & AI & Software Testing & 1995 - 2022\\
Kotti et al.\ \cite{kotti2023machine} & ML & Software Engineering & 2009 - 2022\\
\textbf{Our work} & \textbf{LLM} & \textbf{CGTs} & \textbf{2017 - 2025}\\
\bottomrule
\end{tabular}}
\end{table}

\section{Systematic Review Methodology}
\label{sec:methodology}
To conduct a tertiary SLR, we adopt Kitchenham’s systematic review methodology (for software engineering) \cite{kitchenham2007guidelines} and their later SEGRESS guidelines (for structured, transparent reporting) \cite{kitchenham2022segress}. In line with these guidelines we make an SLR protocol and data collection records publicly available for transparency and reproducibility \cite{zenodo}.

\subsection{Research Questions}
Following Kitchenham’s SLR methodology, the research questions (RQs) are formulated as:

\begin{enumerate}  
\item \textbf{RQ1}: What is the landscape of secondary studies that report (at least in part) on LLM-based CGTs?
Sub-questions:
\begin{enumerate}
    \item What is the distribution of secondary studies by year and by study type (e.g. mapping studies/systematic reviews)?
    \item How many primary studies are reviewed in these secondary studies?
    \item In what venues are they published?
    \item What is AI and SE domain scopes of these studies?
\end{enumerate}
\item \textbf{RQ2}: What HELM measures of LLM-based CGTs are reported in secondary studies? 
\item \textbf{RQ3}: What HELM scenarios of LLM-based CGTs are reported in secondary studies?
\item \textbf{RQ4}: What challenges in applying or integrating LLM-based CGTs into software engineering workflows are identified in secondary studies?
\item \textbf{RQ5}: What future directions for LLM-based CGTs are reported on in existing secondary studies?
\end{enumerate}

\subsection{Search Strategy}
The search process has the following major steps:
\begin{enumerate}
    \item[0] (Preliminary) Search string construction;
    \item[1] Search in digital databases and screening of relevant papers to produce a set of 'seed' papers;
    \item[2] Backward and forward snowballing, starting with 'seed' papers, following screening of newly found (if any) papers;
    \item[3] Full-text screening of papers from step 2 and 3 to identify the final set of secondary studies for review.
\end{enumerate}

\begin{figure}[h]
  \centering
  \includegraphics[width=0.8\textwidth]{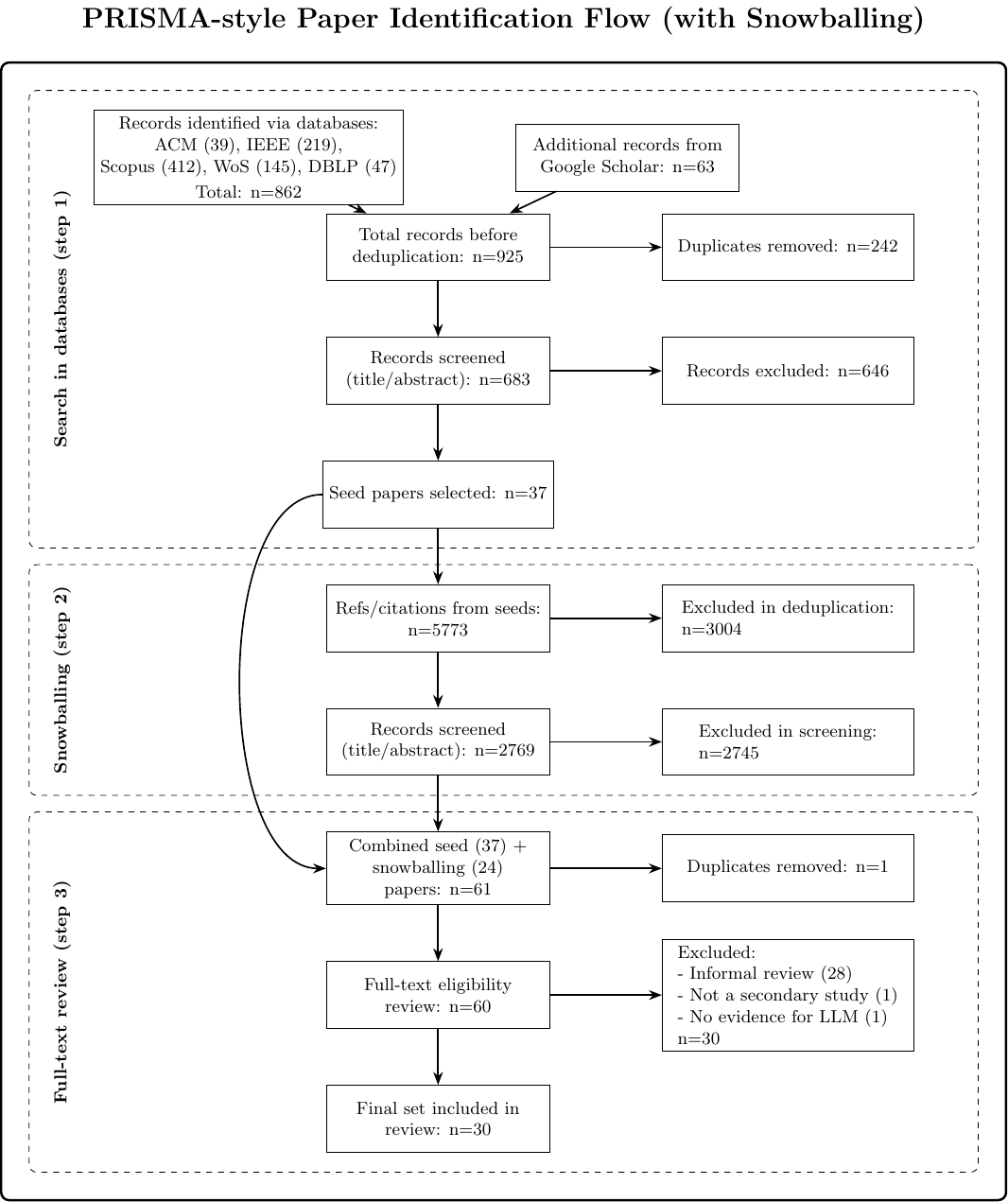}
  \caption{PRISMA flowchart of the tertiary review process.}
  \Description{PRISMA flowchart of the tertiary review process.}
  \label{fig:prisma-flowchart}
\end{figure}

The process is presented in more detail for steps 1, 2, and 3 (omitting the preliminary step that is discussed separately) in Figure~\ref{fig:prisma-flowchart} that follows a PRISMA-like \cite{moher2015preferred} format. 

\subsection{Search String Construction}
Constructing queries for searching in digital publication databases can be challenging. Several approaches like Quasi-Gold Standard (QGS) were proposed to construct such queries in a rigid un-biased manner \cite{kitchenham2013systematic}. QGS can be sensitive to the proper selection of seed publication venues and so here, we used a different approach that employed Google Scholar (GS) to find relevant seed papers. We extracted keywords from these papers for query construction, and then tested the constructed query on a test set of papers (the protocol for this can be found in \cite{zenodo}).

Initially, we searched GS using a trial query constructed from domain-relevant concepts that were agreed by all the authors of this paper. GS was selected due to its ability to fetch significant amount of relevant papers from multiple sources \cite{yasin2020using}. The trial query was constructed drawing from PICOC-style (Population, Intervention, Comparison, Outcome, and Context) query format proposed by Petersen et al. \cite{petersen2015guidelines} and this application is shown below in its abstract form (Comparison and Outcome not being so relevant for mapping studies):

\begin{center}
\begin{minipage}{0.8\linewidth}
\begin{verbatim}

Population: ("code generation" OR "code completion")
Intervention: AND ("LLM" OR "LCM" OR "large language model" OR "large code model")
Context: AND ("systematic review" OR "meta-analysis" OR "survey")

\end{verbatim}    
\end{minipage}
\end{center}

Following this, we looked at the top 20 results returned by GS after running the trial query. All three authors independently screened these top 20 results, looking at abstracts/titles and, if needed, full-text. Each author was tasked with:
\begin{itemize}
    \item Identifying if a paper is relevant based on the inclusion/exclusion criteria provided in Section~\ref{subsec:incl_excl}. If it did not meet the criteria, the reviewer was asked to specify which inclusion or exclusion criteria were not satisfied (e.g. IC1 or EC3, - see Section~\ref{subsec:incl_excl}).
    \item Suggesting additional keywords based on the title/abstract/keywords only to expand the initial trial query. They were asked to structure them into these 3 categories of our PICOC type query: Population, Intervention, and Context.
\end{itemize}

All the authors have discussed their set of reviewed papers reaching agreement on 18/20 papers (where 12/18 were found relevant and 6/18 were found irrelevant). The inter-rater agreement was calculated using Fleiss' kappa \cite{fleiss1971measuring} (suitable when more than 2 raters are involved) and was 0.85, signaling almost perfect agreement.

Next, we randomly split the collected relevant papers into a training set (10/12, used for query construction via aggregated suggested keywords) and a testing set (2/12, held out for validation) using an automated script. Given the small size of the seed set, this split was not intended for statistical performance estimation, but rather as a sanity check to ensure that the constructed query could retrieve relevant studies not explicitly used during query formulation, which is a common practice in query validation for small samples \cite{kohavi1995study}.

The final set of query keywords was agreed after discussion by all the authors of the paper. The resultant final abstract search query is shown below (when searching in databases this query was adapted to best leverage the databases' query format, but preserving semantics):
\begin{center}
\begin{minipage}{0.8\linewidth}
\begin{verbatim}

Population: ("code generation" OR "code completion" OR "software development" OR
 "coding process" OR "programming code process" OR "code generation and review" OR
 "software development lifecycle" OR "SE tasks" OR "program repair" OR "coding" OR
 "debugging" OR "testing" OR "programming" OR "code development" OR "SDLC" OR
 "LLM4SE" OR "generating codes" OR "ai-assisted code generation" OR "generated code")
Intervention: AND ("LLM" OR "LCM" OR "large language model" OR "large code model" OR
     "transformer" OR "Encoder-Decoder")
Context: AND ("systematic review" OR "meta-analysis" OR "survey" OR
     "Systematic Literature Review" OR "Systematic Mapping Study" OR
     "structured methodology" OR "SLR" OR "Mapping Study" OR "literature review")

\end{verbatim}

\end{minipage}
\end{center}
The query was formulated at the level of broader software engineering and code-related activities rather than enumerating individual code-generating tasks. While specific CGTs such as refactoring or code translation are included in our operational definition, pilot searches and prior secondary studies indicate that such tasks are rarely used as primary indexing terms in review titles or abstracts, and are instead discussed under broader categories such as code generation, program repair, or software development. Consequently, we prioritized concept-level terms to maximize recall and reduce sensitivity to heterogeneous task nomenclature. Relevant secondary studies addressing refactoring or translation were subsequently captured during screening and snowballing.

Subsequent validation of this query on the testing set, showed 100\% recall.

\subsection{Inclusion/Exclusion Criteria}
\label{subsec:incl_excl}
For the goals of this SLR, we will use these inclusion criteria (IC) when selecting studies:
\begin{enumerate}
    \item[(IC1)] Secondary studies: SLRs, systematic mapping studies (SMSs), literature surveys, and meta-analyses \cite{kitchenham2007guidelines};
    \item[(IC2)] Studies that focus on, or include, CGTs implemented using large language models (LLMs), as defined in Section~\ref{sec:def-cgts};
    \item[(IC3)] Studies published in peer-reviewed venues and in arxiv (the quality of the latter to be assessed in quality assurance) starting from 2017, inclusive;  
    \item[(IC4)] Studies written in English language.  
\end{enumerate}

We will use these exclusion criteria (EC) to reject studies:  
\begin{enumerate}
    \item[(EC1)] Primary studies that do not summarize existing research;  
    \item[(EC2)] Studies focusing on applying LLMs in software engineering outside of CGTs (e.g., code summarization, code comments, commit messages).  
    \item[(EC3)] Grey literature, except arXiv (we include arXiv due to fast pacing in this research area where many articles appear on arXiv first and the majority is later published in peer-reviewed venues\cite{lin2020many});
    \item[(EC4)] Studies focusing on small scale snippet-oriented LLM generated code for teaching or research activities;
    \item[(EC5)] Studies focusing purely on theoretical AI without software applications or on outdated pre-Transformer models (even if these are large);
    \item[(EC6)] Unavailable.
\end{enumerate}

\subsection{Search Databases}
\label{subsec:libraries}
The automated search was conducted in the following digital libraries, as suggested for systematic literature reviews \cite{gusenbauer2020academic} and commonly used in software engineering secondary studies:
\begin{itemize}
    \item IEEE Xplore \cite{IEEEXplore}  
    \item ACM Digital Library \cite{ACMDL} 
    \item Scopus/ScienceDirect \cite{Scopus, ScienceDirect}
    \item DBLP \cite{DBLP}
    \item Web of Science (WoS) \cite{WebOfScience}
    \item Google Scholar (GS) (as a search engine for additional grey literature like arxiv) \cite{GoogleScholar}
\end{itemize}

\subsection{Study Selection Process}

\subsubsection{Search in digital databases and screening of seed papers}
The search was conducted by the first author of this paper (MC) in the digital libraries (see Section~\ref{subsec:libraries}), adjusting the search string as required for each database. A total of 862 records was returned (excluding those returned in GS), as can be seen in Figure~\ref{fig:prisma-flowchart}.

Given that GS retrieves results using a proprietary relevance-based ranking algorithm and does not support reproducible boolean querying or bulk export, we applied a structured yet practical approach to include it in our review protocol. Following established best practices \cite{gusenbauer2020academic, haddaway2015role}, we screened results in descending order of relevance. Screening was conducted in fixed blocks of 50 results, with a stop criterion defined as follows:
\begin{itemize}
    \item Stop screening if the inclusion rate falls below 5\% across two consecutive 50-result blocks.
    \item Impose a hard cap of 300 results per query.
\end{itemize}

This strategy aligns with literature suggesting that GS’s precision sharply declines beyond the first 200–300 results, and ensures a balance between sensitivity and feasibility \cite{bates2017will, walters2009google}. A total of 63 records were returned and therefore, a total of 925 records (862 from digital databases and 63 from GS) were collected in June 2025. After removing duplicates, 683 were retained for screening.

Following this, the first author (MC) screened 683 records based on their titles and abstracts, and full-text only if needed (if a yes/no decision could not be made on the title/abstract alone). The author marked the paper as yes/no and, if no, specified the IC/EC violated. To assess the reliability of our inclusion/exclusion decisions and to ensure transparency and lack of bias in the screening process, we conducted an inter-rater agreement analysis using Fleiss’ Kappa \cite{fleiss1971measuring}. From the total screening pool of 683 records, we selected a random sample of 30 papers, comprising an equal distribution of 15 included and 15 excluded items (as marked by MC). The two other authors (JB and ME) had to mark this anonymously as yes/no. This balanced sampling design helps mitigate prevalence bias, which can distort kappa estimates in imbalanced datasets \cite{feinstein1990high, mchugh2012interrater}. Methodological sources suggest that a sample size of 30 items is sufficient for reliable estimation of Fleiss’ Kappa, particularly when three or more raters are involved \cite{gwet2014handbook, bujang2017guidelines}. We therefore consider this a sufficiently methodologically-sound compromise between statistical rigor and reviewer workload. Fleiss’ Kappa yielded a value of 0.86, which indicates almost perfect agreement among reviewers \cite{landis1977measurement}. Ultimately, 37 'seed' papers were retained after this step.

\subsubsection{Backward and forward Snowballing}
Snowballing in a literature search is a method used to identify additional relevant papers by exploring the references and citations of a key paper or set of papers (37 'seed' papers in our case) \cite{wohlin2014guidelines}. 

In particular, we employed an iterative snowballing process, following the structured guidelines of Wohlin \cite{wohlin2014guidelines}. To preempt excessive effort during iterations with diminishing returns, we monitored the inclusion rate across snowballing rounds. Specifically, if an iteration yielded fewer than 5\% newly included studies relative to the number of screened references, the process would be terminated early. This threshold is motivated by practical screening guidelines in systematic review literature \cite{gusenbauer2020academic}, which note that the marginal value of additional screening falls sharply below this level. While these studies do not explicitly define 5\% as a stopping point, they acknowledge that precision commonly drops below this threshold in later phases, indicating low yield and high effort. Therefore, we adopt 5\% as a cut-off threshold.


In our study, the yield-based stopping criteria was met after the first iteration where only 24 of 2769 records (0.9\%) screened were deemed relevant see Figure~\ref{fig:prisma-flowchart}.

To manage the screening of 2,769 candidate publications after duplicates were removed, we adopted a hybrid screening approach that combined GPT-4o \cite{openai}, a novel large language model, with manual validation. This is in-line with recommendations in the evidence synthesis literature on semi-automated screening \cite{wallace2010semi,o2015using}, where the AI system is used as a first-pass screener combined with human evaluation of AI decisions. The model was prompted using our inclusion and exclusion criteria (see Section~\ref{subsec:incl_excl}) and tasked with classifying all records based on their titles/abstracts into ``Include'', ``Uncertain'', or ``Exclude'' categories (see the supplementary material for the prompts \cite{zenodo}). 

To evaluate the AI system's reliability, the first author (MC) conducted a validation study on a sample of 100 records drawn from the full corpus of 2,769. The size of a sample aligns with best practices in the literature where 100-200 sample size is commonly used to validate human-machine decisions \cite{waffenschmidt2019single, van2020asreview}. It is also consistent with recommendations for Cohen's kappa estimation (inter-rater agreement), where over 50 samples are suggested \cite{sim2005kappa}. The sample comprised all 31 records that the AI system flagged as relevant (“Include” only as there were no "Uncertain"), together with 69 randomly selected records that the system excluded (“Exclude”).

The first author independently screened this validation set. Among the 31 AI-flagged records, 24 were confirmed relevant (precision = 0.77; Wilson 95\% CI (confidence interval): 0.60–0.89). All 24 human-identified relevant records were captured by the AI (recall = 1.00; Wilson 95\% CI: 0.86–1.00). In the 69 randomly sampled records marked by AI system as "Exclude", none were found relevant (0/69; one-sided exact 95\% upper bound on prevalence = 4.25\%). Agreement between human and AI screening decisions was almost perfect (Cohen’s kappa = 0.83).

\subsubsection{Selection for full-text review}
The final set of 37 'seed' papers and 24 papers identified during the snowballing, resulted in 61 in total, and they were selected for full-text review. During this review by the first author (MC):
\begin{itemize}
    \item 28 were removed because they lacked a systematic approach to literature review and were informal reviews;
    \item one was a duplicate;
    \item one was found to not be a secondary study, upon closer inspection;
    \item one did not provide any evidence that LLMs were reviewed as part of deep learning.
\end{itemize}

A final, total set of 30 papers was selected for our tertiary study (see Figure~\ref{fig:prisma-flowchart}).

\subsection{Quality Assessment}
Each selected study was evaluated by the first author using the DARE (Database of Abstracts of Reviews of Effects) criteria, following the questions and scoring criteria proposed by Kitchenham et al. \cite{kitchenham2010systematic}. There are 4 questions, where each question can score either 1/0.5/0 (for exact questions and scoring see Section 2.4 in Kitchenham et al.\cite{kitchenham2010systematic}. We also include these in our protocol in supplementary material \cite{zenodo}). Reliability was assessed by the other two authors using the same methodology and reviewing 5 different papers each.

\begin{table}
\caption{Quality assessment scores}
\label{tab:qa_scores}
\resizebox{0.7\textwidth}{!}{\begin{tabular}{llccccccc}
\toprule
S & Authors & Q1 & Q2 & Q3 & Q4 & Total & Quality & Reference\\
\midrule
S1  & Vitale et al.    & 1   & 1   & 0 & 1   & 3   & High   & \cite{vitale2025catalog}\\
S2  & Dong et al.      & 0.5 & 0.5 & 0 & 0.5 & 1.5 & Low    & \cite{dong2024pilot}\\
S3  & Liu et al.       & 0   & 0.5 & 0 & 0.5 & 1   & Low    & \cite{liu2024pilot}\\
S4  & Zheng et al.     & 0.5 & 0.5 & 0 & 0.5 & 1.5 & Low    & \cite{zheng2023survey}\\
S5  & Zhang et al.     & 0   & 0.5 & 0 & 0.5 & 1   & Low    & \cite{zhang2023survey}\\
S6  & Gorissen et al.  & 0.5 & 0.5 & 0 & 0.5 & 1.5 & Low    & \cite{gorissen2024survey}\\
S7  & Huang et al.     & 0.5 & 0.5 & 0 & 1   & 2   & Medium & \cite{huang2023survey}\\
S8  & Jiang et al.     & 1   & 1   & 1 & 1   & 4   & High   & \cite{jiang2024survey}\\
S9  & Zhang et al.     & 0.5 & 0.5 & 0 & 1   & 2   & Medium & \cite{zhang2023survey2}\\
S10 & Joel et al.      & 1   & 1   & 0 & 1   & 3   & High   & \cite{joel2024survey}\\
S11 & Ramirez-Rueda et al. & 0.5 & 1 & 1 & 1   & 3.5 & High  & \cite{ramirez2024systematic}\\
S12 & Zhang et al.     & 1   & 0.5 & 0 & 0.5 & 2   & Medium & \cite{zhang2024systematic}\\
S13 & Hu et al.        & 1   & 1   & 0 & 1   & 3   & High   & \cite{hu2025assessing}\\
S14 & Dehaerne et al.  & 1   & 0   & 0 & 1   & 2   & Medium & \cite{dehaerne2022code}\\
S15 & Huangzhao et al. & 0   & 0   & 0 & 0   & 0   & Low    & \cite{huangzhao2024deep}\\
S16 & Wan et al.       & 0   & 0.5 & 0 & 0.5 & 1   & Low    & \cite{wan2024deep}\\
S17 & Lee et al.       & 1   & 0   & 0 & 0.5 & 1.5 & Low    & \cite{lee2025hallucination}\\
S18 & Zhou et al.      & 1   & 1   & 1 & 1   & 4   & High   & \cite{zhou2025large}\\
S19 & Husein et al.    & 1   & 1   & 1 & 1   & 4   & High   & \cite{husein2025large}\\
S20 & Hou et al.       & 1   & 1   & 1 & 1   & 4   & High   & \cite{hou2024large}\\
S21 & G{\"o}rmez et al.& 1   & 0.5 & 0 & 0.5 & 2   & Medium & \cite{gormez2024large}\\
S22 & Utomo et al.     & 1   & 0.5 & 0 & 1   & 2.5 & Medium & \cite{utomo2024machine}\\
S23 & She et al.       & 1   & 1   & 1 & 0.5 & 3.5 & High   & \cite{she2023pitfalls}\\
S24 & Tony et al.      & 1   & 0.5 & 0 & 1   & 2.5 & Medium & \cite{tony2024prompting}\\
S25 & Wang et al.      & 1   & 1   & 1 & 1   & 4   & High   & \cite{wang2024software}\\
S26 & Ahmed et al.     & 1   & 1   & 0 & 1   & 3   & High   & \cite{ahmed2023source}\\
S27 & Ram{\'\i}rez et al.& 1 & 1   & 0 & 1   & 3   & High   & \cite{ramirez2024state}\\
S28 & Sasaki et al.    & 0   & 0.5 & 0 & 1   & 1.5 & Low    & \cite{sasaki2024systematic}\\
S29 & Zubair et al.    & 1   & 1   & 1 & 1   & 4   & High   & \cite{zubair2025use}\\
S30 & Zheng et al.     & 0.5 & 1   & 0 & 0.5 & 2   & Medium & \cite{zheng2025towards}\\
\bottomrule
\end{tabular}
}
\end{table}

The resultant quality assessment (QA) scores can be seen in Table~\ref{tab:qa_scores} where S is a paper identifier, followed by 'Authors', Q is a question with a score assigned, followed by 'Total', 'Quality' is a paper placement into quality bands based n its score, and followed by reference to this paper. The average quality score is 2.45 comparable to other tertiary reviews such as Kitchenham's et al. \cite{kitchenham2010systematic} where average scores of 2-3 are reported. The quality was assessed to guide synthesis: we did not remove papers based on their scores. Instead, we retained all papers for descriptive mapping (e.g., trends, taxonomy) and note their quality when synthesizing data and drawing conclusions. For this reason, papers were partitioned into a Low [0–<2], Medium [2–<3], and High [3–4] categories according to their total QA scores. These cut-points align with generic quality conventions (low/moderate/high) \cite{amalfitano2023artificial, kotti2023machine}. Overall, the distribution of quality categories was Low 9/30 (30.0\%), Medium 8/30 (26.7\%), High 13/30 (43.3\%) with high-medium quality paper prevalence (70\%). 

For bias assessment, inter-rater reliability on score overlaps was estimated using quadratic-weighted Cohen’s k for each question score and for the total. Agreement was almost perfect (exceeding 0.81 \cite{landis1977measurement}): for MC–JB on the total (n=5, k=0.91) and for MC–ME on the total (n=5, k=0.82). Category-level (low/medium/high) classifications were identical for all MC–JB overlaps (5/5) and for 4/5 MC–ME overlaps (the remaining case differed by 0.5 on the total), yielding substantial agreement overall. Subsequently, discrepancies were discussed, and several of MC’s assessments were adopted after consensus, reflecting convergence toward a shared understanding of the criteria. Given this calibration and the observed high agreement level, MC was deemed a reliable assessor to complete the remaining QA evaluations independently.


\subsection{Data Extraction \& Synthesis}
Data extraction was completed by the first author (MC) and cross-checked by the two other authors each validating a randomly assigned set of 5 papers.

\subsubsection{Data Extraction Fields}

\begin{table}[ht]
\centering
\caption{Data extraction fields and their in-paper location}
\label{tbl:sec3-data-rq1}
\resizebox{\textwidth}{!}{

\begin{tabular}{@{}llrlc@{}}
\toprule
\textbf{Data field name} & \textbf{Description} & \textbf{Cardinality} & \textbf{In-paper data location} & \textbf{Coding} \\
\midrule
\multicolumn{5}{l}{\textbf{RQ1}}\\
\midrule
Year of publication & The publication year of the paper & [1] & Metadata & N \\
AI scope & The AI scope as reported in the paper & [1] & Title/Abstract & Y \\
SE KA/Task & Reported SE KA (according to SWEBOK) or CGT & [0..n] & Full-text & Y (SE scope) \\
Venue name & The name of the venue where a paper was published & [1] & Metadata & N \\
Venue type & The type of the venue where a paper was published & [1] & Metadata & N \\
Venue ranking & Ranking of the venue where the paper was published & [0-1] & External (CORE/Scimago) & N \\
\# primary studies & The number of primary studies reviewed in a paper & [1] & Full-text & N \\
Study type & The type of secondary study as reported by the authors & [1] & Full-text & N \\
\addlinespace
\multicolumn{5}{l}{\textbf{RQ2}}\\
\midrule
HELM measure & Synthesized effects reported for LLM-based CGTs & [0..n] & Results/Discussion/Findings & Y\\
&&&and similar sections & \\
\addlinespace
\multicolumn{5}{l}{\textbf{RQ3}}\\
\midrule
\textit{HELM scenario} &&&&\\
Task & CGT as reported in the paper & [0..n] & Full-text & N \\
Language & Programming language(s) as reported in the paper & [0..n] & Full-text & N \\
Application Type & Application type(s) as reported in the paper & [0..n] & Full-text & N \\
Application Domain & Application domain(s) as reported in the paper & [0..n] & Full-text & N \\
\addlinespace
\multicolumn{5}{l}{\textbf{RQ4}}\\
\midrule
Challenges & Synthesized challenges reported for LLM-based CGTs & [0..n] & Results/Discussion/ & Y\\
&&&Implications/Guidelines &  \\
\addlinespace
\multicolumn{5}{l}{\textbf{RQ5}}\\
\midrule
Future directions & Future direction(s) as reported in the papers & [0..n] & Conclusion/Future Work/ & Y\\
&&&Limitations/Discussion & \\
\bottomrule
\end{tabular}
}
\end{table}

For each included study we extracted the data using the extraction form fields presented in Table~\ref{tbl:sec3-data-rq1}. The table shows data field names of the data, its description, cardinality, where in the paper this data was looked for, and if it needs further coding (e.g. heterogeneous / qualitative data). The cardinality specifies how many values could be extracted for a data field. For example, "Year of publication" can only have one value $[1]$, "SE KA/Task" can have zero or more $[0..n]$, and "Venue ranking" may or may not have a value $[0-1]$. In-paper data location specifies where in the paper we looked for the data. In addition to in-paper data, we extracted supplementary metadata (such as "Year of publication", for example) and external data (such as "Venue ranking"). Specifically for the latter, we retrieved venue rankings from CORE \cite{corePortal} (for conferences) and Scimago \cite{scimagoJR}(for journals). Here values were not coded but directly recorded as provided by the external source. For all non-metadata fields, we have recorded the exact location of the identified data (for example, section, page numbers, figures or tables). AI scope was recorded as provided by the authors and cross-checked with standard definitions from AI/ML/DL/LLM taxonomy derived from Russell and Norvig \cite{russell2021artificial}. (All AI scope extracted data matched those in the taxonomy). The "Study type" was recorded as provided by the authors. Finally, when extracting evidence for "HELM measure", "Challenges", and "Future directions" we followed these rules:
\begin{itemize}
    \item We collected synthesized evidence only: Evidence had to be from across multiple techniques, approaches, or studies, providing multiple references for a claim, and summarizing multiple such sources.
    \item We only collected evidence related to LLMs and CGTs (in case when papers had broader AI/SE scope). 
    \item We recorded verbatim exact quotes extracted from the papers in support of the effects, challenges and directions (including their location in the paper, as stated earlier).
\end{itemize}

\subsubsection{Data coding}
For items flagged ``Y'' in the "Coding" column of Table~\ref{tbl:sec3-data-rq1}, we apply predefined schemes to normalize heterogeneous terminology.

AI scope was coded with standard definitions (for AI/ML/DL/LLM) derived from Russell and Norvig \cite{russell2021artificial}. In the majority of cases, each study was assigned a single AI scope label reflecting its primary methodological paradigm. In one case, a study made substantive and inseparable use of both ML-based and non-AI techniques: this study was therefore coded with a hybrid label (ML \& Non-AI) rather than forcing an artificial dominance decision.

For	"SE KA/Task" (coded name "SE scope, see Table~\ref{tbl:sec3-data-rq1}") SWEBOK v4 KAs were used \cite{SWEBOK2024}. The mapping protocol was as follows:
\begin{enumerate}
    \item First we checked if KAs are explicitly provided (e.g. a paper might structure itself explicitly around KAs) aligning those KAs to SWEBOK v4. For example, some KAs have newer names in SWEBOK v4: e.g. "software development" -> "software construction", "software security" KA now encompasses tasks like vulnerability repair. In this case KAs are reported as they appear in the paper (adjusting to SWEBOK v4, only if needed).
    \item If KAs were not explicitly provided, we tried to match software engineering tasks discussed in a paper to KAs, using KA and task definitions: e.g. "vulnerability repair" -> "software security", "bug repair" -> "software construction/software maintenance".
    \item In other cases, when a match was not apparent, we recorded a "Cannot map".
\end{enumerate}

\begin{table}[t]
\centering
\caption{Challenges categories, their definitions, and their relation to existing taxonomies.}
\label{tbl:rq_challenges}
\resizebox{\textwidth}{!}{\begin{tabular}{p{3.2cm} p{7.4cm} p{5.2cm}}
\toprule
\textbf{Category} & \textbf{Definition} & \textbf{Relation to existing taxonomies}\\
\midrule
Data \& Context & Mismatch between available context and task needs (e.g., private APIs, repo-scale grounding, token limits, domain shift, early-phase data sparsity). & HELM: Robustness, Accuracy; ISO/IEC~25010: Compatibility, Reliability;\\
\addlinespace
Tooling \& Workflow & Integrating LLMs-based CGTs into existing tools or workflows. & ISO/IEC~25010: Interaction Capability;\\
\addlinespace
Evaluation \& Benchmark Validity & Benchmarks/metrics not reflecting practice (ambiguous tasks, single-metric focus, leakage/imbalance, narrow languages/projects). & HELM: Robustness, Bias;\\
\addlinespace
People \& Process & Human/organizational factors (trust issues, review workload, ownership, missing communication). & \\
\addlinespace
Legal \& Privacy & Privacy and intellectual property risks. & HELM: Toxicity; ISO/IEC~25010: Security;\\
\addlinespace
Security \& Safety & Vulnerabilities in generated code; adversarial fragility, data poisoning. & HELM: Toxicity; ISO/IEC~25010: Security;\\
\addlinespace
Economics & Deployment/runtime constraints (latency/throughput, cost, energy, hardware and software limits). & HELM: Efficiency; ISO/IEC~25010: Performance efficiency;\\
\addlinespace
Model/Training & Limits from pretraining/fine-tuning/continual learning and representations (data scarcity, lack of structural information). & HELM: Robustness, Accuracy;\\
\addlinespace
Not provided & For studies that explicitly do not state a challenge: retained for methodological transparency. & \\
\addlinespace
Other/Uncategorized & For statements that do not fit well into other categories. Necessary to ensure no force-fitting, maintaining methodological rigor. & Kitchenham \cite{kitchenham2007guidelines}\\
\bottomrule
\end{tabular}}
\end{table}

\begin{table}[htbp]
\centering
\caption{Future direction categories, their definitions, and their relation to existing taxonomies.}
\label{tab:future_directions_categories}
\resizebox{0.8\textwidth}{!}{
\begin{tabular}{p{3.2cm} p{7.4cm} p{5.2cm}}
\toprule
\textbf{Category} & \textbf{Definition} & \textbf{Relation to existing taxonomies} \\
\midrule
Model improvement \& training & Covers future work aimed at improving model architectures, training strategies, fine-tuning, performance, explainability, and prompting techniques. & Justified by \emph{SWEBOK} (Software Construction) for code-level improvements and HELM measures such as accuracy, robustness, efficiency. ISO/IEC 25010 also emphasizes \emph{performance efficiency}. \cite{iso25010, SWEBOK2024, liang2022holistic}\\
\addlinespace
Benchmark \& evaluation improvement & Includes directions for creating benchmarks, metrics, and multi-level evaluation protocols.  & Justified by \emph{SWEBOK} (Software Testing), HELM metrics (accuracy, calibration, robustness), and ISO/IEC 25010 (\emph{functional suitability} and \emph{quality in use}). \cite{iso25010, SWEBOK2024, liang2022holistic} \\
\addlinespace
Advancing experimental research & Captures the need for more empirical, large-scale, and reproducible studies to strengthen evidence bases. & In line with Amalfitano et al. \cite{amalfitano2023artificial}\\
\addlinespace
Dataset improvement & Refers to improving training/evaluation datasets, including quality, representativeness, coverage, and benchmarks.  & Grounded in \emph{SWEBOK} (Software Construction inputs) and ISO/IEC 25010 (data quality as part of product quality). \cite{SWEBOK2024, iso25010} \\
\addlinespace
Ensembling \& hybridization & Future work directions suggesting combinations of models (e.g., ensembles, hybrid systems with symbolic methods). &  Justified as a standard AI approach and a common pattern in mature research fields, where complementary techniques are combined to mitigate individual limitations \cite{dietterich2000ensemble}. \\
\addlinespace
Integration into practice & Focuses on deploying LLMs into real-world SE workflows, adoption in industry, and socio-technical integration. &  Supported by \emph{SWEBOK} (Software Engineering Management, Software Process) and ISO/IEC 25010 (\emph{maintainability} and \emph{portability}). \cite{SWEBOK2024, iso25010} \\
\addlinespace
Security, Privacy \& Trustworthiness & Consolidates concerns about vulnerabilities, adversarial robustness, privacy, and transparency.  & Justified by HELM (\emph{robustness}, \emph{toxicity}), and ISO/IEC 25010 (\emph{security} as a core characteristic). \cite{liang2022holistic, iso25010}\\
\addlinespace
Not provided & For studies that explicitly do not state a future direction: retained for methodological transparency. & \\
\addlinespace
Other/Uncategorized & For statements that do not fit well into other categories. Necessary to ensure no force-fitting, maintaining methodological rigor. & Kitchenham \cite{kitchenham2007guidelines}\\
\bottomrule
\end{tabular}

}
\end{table}

"HELM measure" data was assigned to HELM measure categories as described in Section~\ref{subsec:helm_framework}. "Challenges" were coded using the categories shown in Table~\ref{tbl:rq_challenges}. We derived the challenge categories inductively, iteratively, in a bottom-up manner using thematic synthesis tailored to software engineering and grounding categories in existing taxonomies, where possible (see Table~\ref{tbl:rq_challenges}). The first author (MC) iteratively performed clustering to consolidate codes into higher-order, non-overlapping categories. Two other authors reviewed these categories over several meetings, and disagreements were reconciled/the categories refined. This choice of thematic synthesis is consistent with established SE guidance for heterogeneous bodies of evidence where meta-analysis is infeasible, provided that extraction forms and procedures are reported \cite{kitchenham2007guidelines, kitchenham2022segress}. Although 'Tooling \& Workflow' issues may indirectly affect efficiency (e.g., developer latency), we treat them separately from HELM’s efficiency dimension, which we reserve for resource and performance-oriented constraints captured under the Economics category. 'People \& Process issues do not map cleanly to existing technical evaluation frameworks such as HELM or ISO/IEC 25010, reflecting a known gap in current taxonomies with respect to human and organizational factors in LLM-based software engineering. Likewise (following the same procedure as for "Challenges"), "Future directions" were assigned to the categories defined in Table~\ref{tab:future_directions_categories}. Here, specifically, we distinguish 'Dataset improvement' and 'Benchmark \& evaluation improvement' analytically: dataset improvement concerns the quality and coverage of data artifacts themselves, whereas benchmark and evaluation improvement focuses on task formulation, metrics, and evaluation protocols applied to those datasets.

The final set of papers along with their references and data extraction items with single values (e.g., "Year of publication", "Study type") is reported in the Table~\ref{tab:final_set}. Items with potentially multiple values (e.g., "SE KA/Task", "HELM measure") are summarized in aggregated form, with the full per-study data available in the supplementary material \cite{zenodo}.

\begin{table}[htbp]
\centering
\caption{Final set of secondary studies with extracted single-valued data and references}
\label{tab:final_set}
\resizebox{\textwidth}{!}{\begin{tabular}{l c c c l l r c r c}
\toprule
\textbf{S} & \textbf{Year} & \textbf{AI scope} & \textbf{Venue name} & \textbf{Venue type} & \textbf{Venue ranking} & \textbf{\# primary} & \textbf{Study type} & \textbf{QA} & \textbf{Reference} \\
& \textbf{of publication} & & (abbreviation) & & & \textbf{studies} &  & \textbf{score} &\\
\midrule
S1  & 2025 & LLM            & TOSEM     & Journal   & Q1                  & 107 & SLR & 3   & \cite{vitale2025catalog} \\
S2  & 2024 & LLM            & SEA4DQ    & Workshop     & N/A                 & 28  & Survey & 1.5 & \cite{dong2024pilot} \\
S3  & 2024 & LLM            & TAAI      & Conference     & N/A                 & 28  & SLR & 1   & \cite{liu2024pilot} \\
S4  & 2023 & LLM            & arXiv     & Preprint archive        & --                  & 134 & Review & 1.5 & \cite{zheng2023survey} \\
S5  & 2023 & DL             & TOSEM     & Journal   & Q1                  & 112 & Survey/SLR & 1   & \cite{zhang2023survey} \\
S6  & 2024 & LLM            & HCSE      & Conference     & N/A                 & 55  & Survey & 1.5 & \cite{gorissen2024survey} \\
S7  & 2023 & ML \& non-AI   & arXiv     & Preprint archive        & --                  & 140 & Survey & 2   & \cite{huang2023survey} \\
S8  & 2024 & LLM            & arXiv     & Preprint archive        & --                  & 235 & Survey/SLR & 4   & \cite{jiang2024survey} \\
S9  & 2023 & LLM            & arXiv     & Preprint archive        & --                  & 1009& Survey & 2   & \cite{zhang2023survey2} \\
S10 & 2024 & LLM            & arXiv     & Preprint archive        & --                  & 111 & Survey/SLR & 3   & \cite{joel2024survey} \\
S11 & 2024 & AI   & PCS       & Journal   & Q4                  & 20  & SLR & 3.5   & \cite{ramirez2024systematic} \\
S12 & 2024 & LLM            & arXiv     & Preprint archive        & --                  & 110 & SLR & 2   & \cite{zhang2024systematic} \\
S13 & 2025 & LLM            & arXiv     & Preprint archive        & --                  & 291 & SLR & 3   & \cite{hu2025assessing} \\
S14 & 2022 & ML             & IEEE Access & Journal & Q1                  & 37  & Review & 2   & \cite{dehaerne2022code} \\
S15 & 2024 & DL             & SCIS      & Journal   & Q1                  & 142 & Survey & 0   & \cite{huangzhao2024deep} \\
S16 & 2024 & DL             & ACM CSUR  & Journal   & Q1                  & 276 & Survey/Review & 1   & \cite{wan2024deep} \\
S17 & 2025 & LLM            & arXiv     & Preprint archive        & --                  & 41  & Survey/SLR & 1.5 & \cite{lee2025hallucination} \\
S18 & 2025 & LLM            & TOSEM     & Journal   & Q1                  & 58  & SLR & 4   & \cite{zhou2025large} \\
S19 & 2025 & LLM            & CSI       & Journal   & Q1                  & 23  & SLR & 4   & \cite{husein2025large} \\
S20 & 2024 & LLM            & TOSEM     & Journal   & Q1                  & 395 & SLR & 4   & \cite{hou2024large} \\
S21 & 2024 & LLM            & EuroSPI   & Conference     & B                   & 7   & SMS & 2   & \cite{gormez2024large} \\
S22 & 2024 & ML             & ICITCOM   & Conference     & N/A                 & 36  & SLR & 2.5 & \cite{utomo2024machine} \\
S23 & 2023 & DL             & TOSEM     & Journal   & Q1                  & 67  & Survey/SLR & 3.5 & \cite{she2023pitfalls} \\
S24 & 2024 & LLM            & arXiv     & Preprint archive        & --                  & 13  & Systematic investigation/SLR & 2.5 & \cite{tony2024prompting} \\
S25 & 2024 & LLM            & TSE       & Journal   & Q1                  & 102 & Review & 4   & \cite{wang2024software} \\
S26 & 2023 & DL             & EPIA      & Conference     & National:Portugal   & 29  & Survey & 3   & \cite{ahmed2023source} \\
S27 & 2024 & LLM            & CONISOFT  & Conference     & N/A                 & 15  & SLR & 3   & \cite{ramirez2024state} \\
S28 & 2024 & LLM            & COMPSAC   & Conference     & B                   & 28  & SLR & 1.5 & \cite{sasaki2024systematic} \\
S29 & 2025 & LLM            & CSI       & Journal   & Q1                  & 41  & SLR & 4   & \cite{zubair2025use} \\
S30 & 2024 & LLM            & ESE       & Journal   & Q1                  & 123 & Review & 2   & \cite{zheng2025towards} \\
\bottomrule
\end{tabular}}
\end{table}

\subsubsection{Data synthesis}
Using the coded dataset, we aggregate and analyze data using Kitchenham's guidelines \cite{kitchenham2007guidelines} to answer the RQs: 
\begin{itemize}
    \item Descriptive summaries (counts, proportions, timelines) for single-valued items (e.g. "Year of publication", "\# primary studies");
    \item Distributions and cross-tabulations for multi-valued items (e.g., "SE KA/Task" coded as "AI scope";
    \item Thematic grouping of qualitative data such as "HELM measures", "Challenges", and "Future directions".    
\end{itemize}

\subsubsection{Risk of Bias Assessment}
\paragraph{AI-scope} Each study was assigned an AI scope label (nominal categories: LLM, DL, ML, ML \& non-AI, AI). MC (the first author) labeled all studies; JB and ME independently labeled pre-specified overlapping subsets (5 papers each, one shared). After the calibration round, near-complete agreement was achieved on the AI-scope field. MC and JB reached almost perfect agreement (using Gwet's AC1 for highly unbalanced categories \cite{gwet2008computing}) (Gwet's AC1 = 1.00), while MC and ME reached substantial agreement (Gwet's AC1 = 0.76). These results confirm that the main extractor’s coding can be considered reliable for the remaining studies.



\paragraph{SE KA/Task} Because each study could receive multiple labels here, we looked at how raters agree on sets of KAs per study. For this we employed Jaccard similarity to measure sets overlap and Krippendorff's $\alpha$ for statistical measure of agreement on sets \cite{krippendorff2018content}. Across the overlapping subset of studies, the mean Jaccard similarity between raters was 0.85 (MC–JB) and 0.97 (MC–ME), showing that the raters selected largely identical sets of categories. When assessed using Krippendorff’s $\alpha$, the coefficient was 0.875, indicating strong, almost-perfect reliability ($\alpha$ > 0.80). 
 

\paragraph{RQ3: scenario} Likewise, all data attributes representing a scenario (i.e, \textit{task, language, application type, and application domain}) can all have multiple values, therefore we need to assess how raters agree on these using Jaccard and Krippendorff's $\alpha$. Here, the mean Jaccard similarity for \textit{task} between raters was 0.88 (MC–JB) and 0.92 (MC–ME), suggesting that the raters selected almost identical sets of tasks. When assessed using Krippendorff’s $\alpha$, the coefficient was 0.84, indicating almost-perfect reliability ($\alpha$ > 0.80). For \textit{language}, Jaccard similarity was 0.8 (MC-JB) and 1.0 (MC-ME) with Krippendorff's $\alpha=0.75$, suggesting very high overlap with substantial reliability. Similarly for \textit{application type} Jaccard was 0.89 (MC-JB) and 1.0 (MC-ME) with	$\alpha=0.77$ indicating very high overlap and substantial reliability. Finally, raters completely agreed on \textit{domain} with Jaccard of 1.0 for both raters. Because there was no variability in coding decisions (all raters consistently selected no applicable categories), Krippendorff’s $\alpha$ cannot be computed: the measure is undefined in the absence of variance.

\paragraph{RQ2, RQ4, and RQ5 extraction reliability} To assess the reliability of our data extraction process, we adopted the \textit{extractor–checker} strategy recommended by \cite{kitchenham2007guidelines}, where one researcher (the extractor - MC) identifies evidence quotes and assigns labels, and two independent checkers (JB and ME) validate random samples of the extracted data. Each checker inspected an independent random sample of 20 entries drawn from the full dataset for each RQ2 RQ4 and RQ5, classifying each extraction as \textit{ACCEPT} (quote and label correct), \textit{RELABEL} (quote correct, label incorrect), or \textit{REJECT} (quote incorrect).

Based on these checker decisions we defined and computed two reliability metrics. The quote correct rate (QCR) quantifies the proportion of correctly extracted quotes (regardless of label) as
\[
\mathrm{QCR} = \frac{A + R_L}{n} = 1 - \frac{R_X}{n},
\]
where \(A\) is the number of \textit{ACCEPT} items, \(R_L\) the number of \textit{RELABEL} items, \(R_X\) the number of \textit{REJECT} items, and \(n\) the total reviewed items. The label accuracy given correct quote (LAcc) measures labeling correctness conditional on a valid quote:
\[
\mathrm{LAcc} = \frac{A}{A + R_L}.
\]

All proportions were subject to 95\% Wilson score confidence intervals (CI) \cite{newcombe1998two}, which provide more accurate coverage for smaller samples.

Across the three research questions, the extractor demonstrated consistently high reliability (see Table~\ref{tab:reliability}). For RQ2, both QCR and LAcc were nearly perfect (QCR = 1.00\,[0.94--1.00]; LAcc = 0.98\,[0.87--1.00]). For RQ4, performance remained excellent (QCR = 1.00\,[0.96--1.00]; LAcc = 0.98\,[0.87--1.00]). For RQ5, reliability was still high though slightly lower (QCR = 1.00\,[0.95--1.00]; LAcc = 0.95\,[0.83--0.99]), suggesting minor increases in label ambiguity rather than extraction errors. All but lower CI bounds exceed 0.80, suggesting almost perfect reliability. These results confirm that the extractor (MC) can be considered highly reliable for the remaining extractions.

\begin{table}[h]
\centering
\caption{Reliability of data extraction across RQ2,4,5 (95\% Wilson CI).}
\label{tab:reliability}
\begin{tabular}{lcccc}
\toprule
\textbf{RQ} & \textbf{Metric} & \textbf{Value} & \textbf{CI Low} & \textbf{CI High} \\
\midrule
RQ2 & QCR  & 1.00 & 0.94 & 1.00 \\
    & LAcc & 0.98 & 0.87 & 1.00 \\
RQ4 & QCR  & 1.00 & 0.96 & 1.00 \\
    & LAcc & 0.98 & 0.87 & 1.00 \\
RQ5 & QCR  & 1.00 & 0.95 & 1.00 \\
    & LAcc & 0.95 & 0.83 & 0.99 \\
\bottomrule
\end{tabular}
\end{table}

\section{Results}
\label{sec:results}
In this section we present and discuss the results of data analysis to answer the RQs of this work.

\subsection{RQ1: What is the landscape of secondary studies that report (at least in part) on LLM-based CGTs?}

\subsubsection{RQ1a: What is the distribution of secondary studies by year and by study type (e.g. mapping studies/systematic
reviews)?}
\label{subsubsec:rq1a}
\begin{figure}[h]
  \centering
  \includegraphics[width=0.7\textwidth]{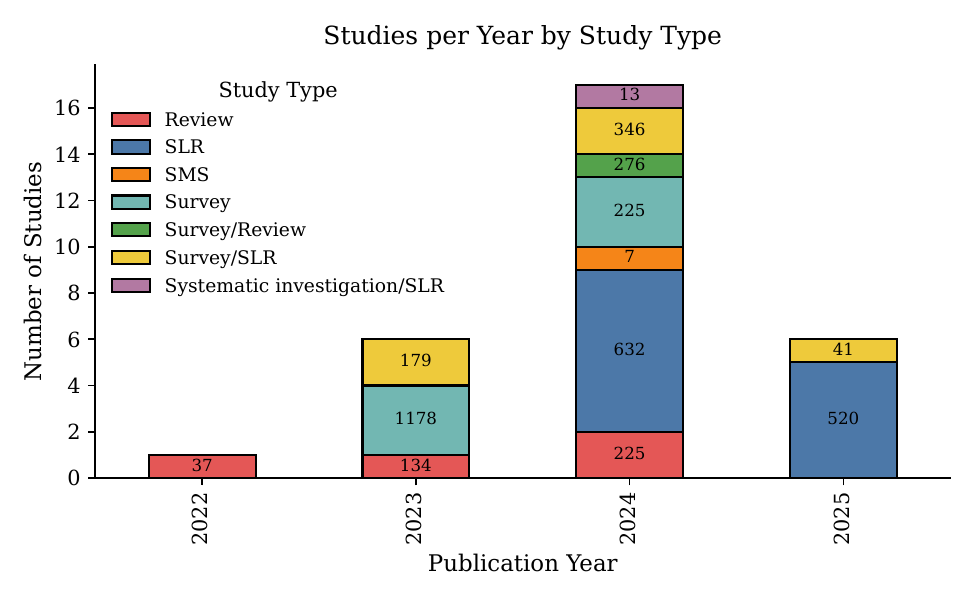}
  \caption{Secondary studies per year by study type}
  \label{fig:study_by_year_type}
  \Description{Secondary studies per year by study type}
\end{figure}
The secondary studies are grouped by the year and by the study type in Figure~\ref{fig:study_by_year_type}. The earliest study in our set of papers dates to 2022, a review published in \emph{IEEE Access} (S14, see Table~\ref{tab:final_set}). In 2023, 1 review, 3 surveys, and 2 survey/SLR appeared, including several in \emph{ACM Transactions on Software Engineering and Methodology (TOSEM)} and \emph{arXiv}, 6 total. The year 2024 marked a notable increase, with 17 studies overall (mostly SLRs and surveys). These studies were widely distributed across venues, including journals such as \emph{TOSEM}, \emph{IEEE Transactions on Software Engineering (TSE)}, and \emph{ACM Computing Surveys (ACM CSUR)}; conferences and workshops such as the \emph{International Workshop on Software Engineering and AI for Data Quality in Cyber-Physical Systems/Internet of Things (SEA4DQ)}, the \emph{International Conference on Technologies and Applications of Artificial Intelligence (TAAI)}, the \emph{International Working Conference on Human-Centered Software Engineering (HCSE)}, and others; and preprint archives (\emph{arXiv}). Finally, by the time the studies were collected in June 2025, we identified six new studies for the year of 2025 (5 SLRs and 1 survey/SLR), appearing in \emph{TOSEM}, \emph{Empirical Software Engineering (ESE)}, and \emph{Computer Standards and Interfaces (CSI)} as well as \emph{arXiv}. Given that the year is incomplete, further growth towards the 2024 "high-tide" mark is expected.

Two noticeable trends can be observed here (see Figure~\ref{fig:study_by_year_type}). First, the data suggests a clear increase of secondary studies including LLM-based CGTs: from a single study in 2022, through 6 studies in 2023, to 17 in 2024, with a leveling off to 6 in the first 6 months of 2025. There is also an observable change in the balance between more exploratory, classification-oriented studies (reviews and surveys) and more directed, synthesis-oriented studies such as SLRs/SMS over time \cite{kitchenham2007guidelines}: 0\% purely SLR/SMS in 2022 and 2023, 47\% in 2024 (8/17), and 83\% in 2025 (5/6). This progression indicates a growing maturity of the research area.

\subsubsection{RQ1b: How many primary studies are reviewed in these secondary studies?}
\label{subsubsec:rq1b}
To answer this RQ, we examined the cumulative number of primary studies (non-unique) reported across the included SLRs and SMSs, as reported in Table~\ref{tab:final_set} and visualized in Figure~\ref{fig:study_by_year_type} (with numbers inside the boxes for each study type and year). As can be seen, in 2022, the single study reviewed only 37 primary studies. By 2023, coverage expanded significantly: six review/survey/early SLR papers collectively reviewed 1491 primary studies that year. In 2024, the number of primary studies increased further: particularly 7 SLRs reviewed 632 primary studies, 1 SMS covered 7, with the rest of primary studies reviewed in surveys/reviews and mixed type studies, resulting in a total of 1724 that year. Noticeable trend here is the growth of the reviewed papers in total. By mid-2025 (our data collection cut-off in June), 5 SLRs had reviewed 520 primary studies and the other survey/SLR covered 41, producing 561 total for the current year; a surprising finding considering the more general growth in AI4SE.

Surveys, reviews and mixed type papers seemed to consistently review a larger number of primary studies than SLRs (e.g. 1092 total in these study types versus 632 total in SLRs for 2024), reflecting their broader and more exploratory aims. This is only different in 2025, but the data for that year is incomplete. Although the number of secondary studies increased nearly threefold from 2023 to 2024 (from 6 to 17), the total number of primary studies reviewed grew less significantly (from 1,491 to 1,724). This trend could suggest both the maturation of the research area (with less primary studies appearing) and the onset of redundancy across secondary studies, with future growth in secondary research likely to stem more from diversification of scope and questions (more SLRs) than from the discovery of entirely new primary studies (less informal reviews and mapping studies). Importantly, these trends are observed over a relatively short time span (approximately 3.5 years) and should therefore be interpreted as early indicators rather than long-term trajectories.

\subsubsection{RQ1c: In what venues are they published?}
\label{subsubsec:rq1c}
\begin{table}
\caption{Distribution of Studies by Venue and Source Type}
\label{tab:venue_summary}
\resizebox{0.6\textwidth}{!}{
\begin{tabular}{llcr}
\toprule
\textbf{Venue name/type} & \textbf{Venue ranking} & \textbf{Count} & \textbf{Average QA score}\\
\midrule
\multicolumn{4}{l}{\textbf{Journal}} \\
\midrule
TOSEM & Q1 & 5 & 3.1 \\
CSI & Q1 & 2 & 4\\
ESE & Q1 & 1 & 2\\
IEEE Access & Q1 & 1 & 2\\
PCS & Q4 & 1 & 3.5\\
SCIS & Q1 & 1 & 0\\
ACM CSUR & Q1 & 1 & 1\\
TSE & Q1 & 1 & 4\\
\midrule
\textbf{Total (Journals)} & & \textbf{13} & \textbf{2.77}\\
\midrule
\multicolumn{4}{l}{\textbf{Conference}} \\
\midrule
COMPSAC & B & 1 & 1.5 \\
CONISOFT & N/A & 1 & 3 \\
EPIA & National:Portugal & 1 & 3\\
EuroSPI & B & 1 & 2\\
HCSE & N/A & 1 & 1.5\\
ICITCOM & N/A & 1 & 2.5\\
TAAI & N/A & 1 & 1\\
\midrule
\textbf{Total (Conferences)} & & \textbf{7} & \textbf{2.07}\\
\midrule
\multicolumn{4}{l}{\textbf{Workshop}} \\
\midrule
SEA4DQ & N/A & 1 & 1.5\\
\midrule
\textbf{Total (Workshop)} & & \textbf{1} & \textbf{1.5}\\
\midrule
\multicolumn{4}{l}{\textbf{Preprint archive}} \\
\midrule
arXiv & --- & 9 & 2.39\\
\midrule
\textbf{Total (Preprint archive)} & & \textbf{9} & \textbf{2.39}\\
\bottomrule
\end{tabular}
}
\end{table}

As can be seen in Table~\ref{tab:venue_summary}, the majority of peer-reviewed secondary studies have been published in journals (13 out of 30). These include 8 venues such as \emph{ACM Transactions on Software Engineering and Methodology (TOSEM)} having 5 studies and \emph{Computer Standards and Interfaces (CSI)} having 2 studies. All other venues had one paper. Most of these journals (7/8) are ranked Q1 in Scimago, suggesting that the topic has gained recognition in high-quality venues. One study was published in PCS, which has Q4 ranking in Scimago, however, the QA score of that study was high (4). 

Conferences account for 7 studies, spanning diverse venues such as \emph{COMPSAC}, and \emph{HCSE}. For 4 of these venues rankings could not be obtained and the remaining 3 had variable rankings (B  and national-level). Only one workshop paper was identified, where a ranking was not available and a low QA score was derived (1.5). This is consistent with the more exploratory, early-stage nature of workshop publications.  

Finally, a substantial number of studies (9 out of 30) appeared on the \textbf{arXiv} preprint server. This reflects the fast-moving nature of the field and a drive for rapid dissemination of survey results. Ranking scores are not available for preprints since they bypass peer review.

Unsurprisingly, the average QA score (2.77) was the highest for studies published in journals, reflecting their top-tier nature and rigorous peer-review. Interestingly though, the second highest average QA score (2.39) was recorded for arXiv preprints. This suggests that, despite the lack of peer-review, some of these studies followed a rigorous, systematic literature review approach. Studies published in conferences had comparable but slightly lower average QA score of 2.07 and the workshop paper had the lowest of 1.5.


\subsubsection{RQ1d: What is AI and SE domain scopes of these studies?}
\label{subsubsec:rq1d}
\begin{table}[htbp]
\centering
\caption{Mapping of papers to AI domain scope.}
\label{tab:papers_by_category_ai}
\resizebox{\textwidth}{!}{
\begin{tabular}{p{5cm}p{1.5cm}p{8.5cm}}
\toprule
\textbf{AI scope} & \textbf{Count} & \textbf{Papers (S identifiers)} \\
\midrule
LLM & 21 & S1, S2, S3, S4, S6, S8, S9, S10, S12, S13, S17, S18, S19, S20, S21, S24, S25, S27, S28, S29, S30 \\
DL & 5 & S5, S15, S16, S23, S26 \\
ML & 2 & S14, S22 \\
ML \& non-AI & 1 & S7 \\
AI & 1 & S11\\
\bottomrule
\end{tabular}}
\end{table}

\begin{table}[htbp]
\centering
\caption{Mapping of papers to SE (SWEBOK) domain scope.}
\label{tab:papers_by_domain_scope}
\resizebox{\textwidth}{!}{
\begin{tabular}{p{5cm}p{1.5cm}p{8.5cm}}
\toprule
\textbf{SE Scope} & \textbf{Count} & \textbf{Papers (S identifiers)} \\
\midrule
Software Construction & 25 & S1, S2, S3, S4, S5, S6, S7, S8, S9, S10, S11, S12, S13, S14, S15, S16, S19, S20, S21, S22, S24, S25, S26, S29, S30 \\
Software Maintenance & 14 & S1, S4, S5, S7, S9, S12, S13, S14, S16, S20, S21, S25, S29, S30 \\
Software Security & 11 & S1, S4, S5, S7, S9, S12, S13, S18, S20, S27, S30 \\
Software Testing & 7 & S2, S4, S9, S13, S20, S21, S25 \\
Software Design & 5 & S2, S9, S13, S20, S21 \\
Software Requirements & 5 & S2, S9, S13, S20, S21 \\
Software Engineering Management & 3 & S9, S20, S21 \\
Software Quality & 2 & S13, S20 \\
Software Configuration Management & 1 & S30 \\
Software Engineering Operations & 1 & S13 \\
\midrule
Cannot Map & 3 & S17, S23, S28 \\
\bottomrule
\end{tabular}}
\end{table}

\begin{table}[h]
\centering
\caption{Distribution of secondary studies by number of SE scope topics (SWEBOK KAs) covered, excluding ``Cannot Map''.}
\label{tab:topics-distribution-clean}
\resizebox{0.75\textwidth}{!}{\begin{tabular}{c c p{9cm}}
\toprule
\textbf{\# SE KAs} & \textbf{Number of Papers} & \textbf{Papers (S identifiers)} \\
\midrule
1 & 12 & S3, S6, S8, S10, S11, S15, S18, S19, S22, S24, S26, S27 \\
2 & 3  & S14, S16, S29 \\
3 & 5  & S1, S5, S7, S12, S25 \\
4 & 3  & S2, S4, S30 \\
6 & 1  & S21 \\
7 & 1  & S9 \\
8 & 2  & S13, S20 \\
\bottomrule
\end{tabular}}
\end{table}

\begin{figure}[h]
  \centering
  \includegraphics[width=\textwidth]{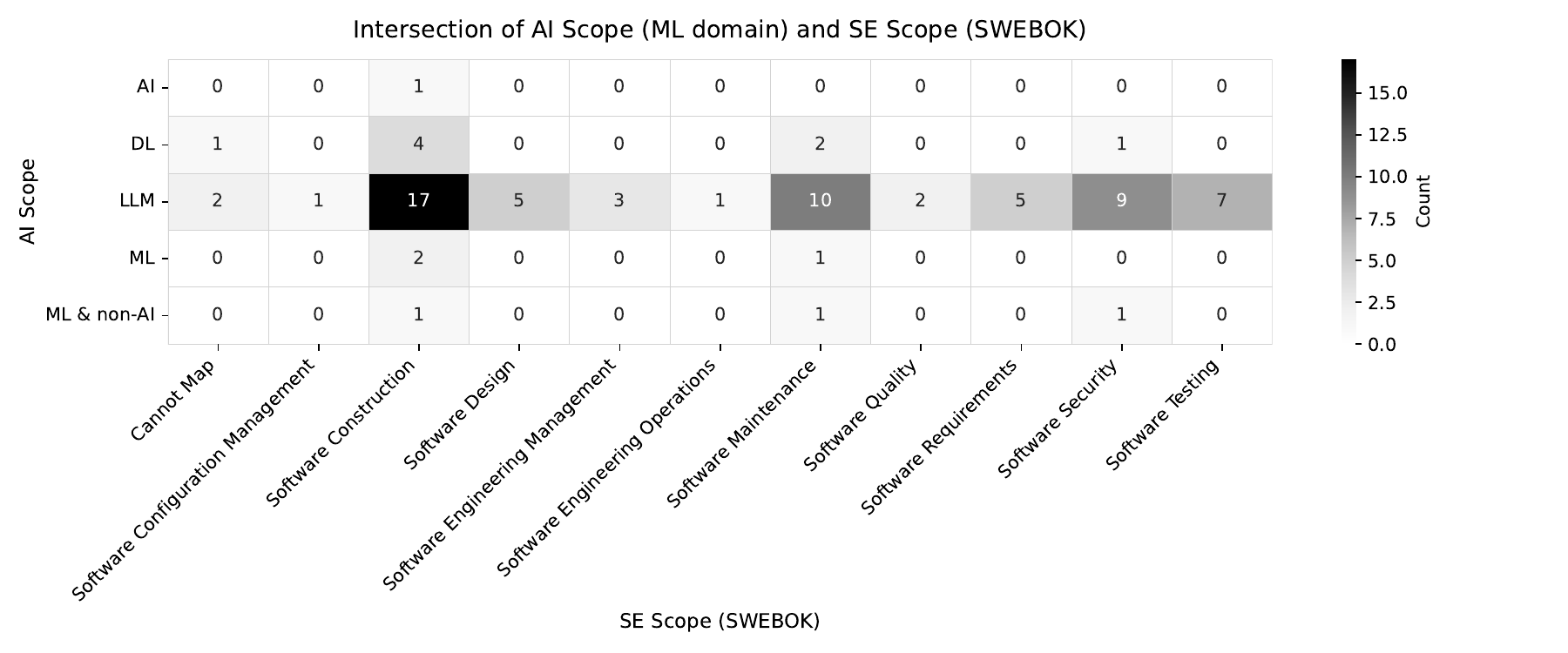}
  \caption{Intersection of AI scope and SE scope}
  \label{fig:ai_se_intersection}
  \Description{Intersection of AI scope and SE scope}
\end{figure}

To address this RQ, we used the AI scope of each secondary study and SE scope (coded) where the latter can have multiple KAs (according to SWEBOK \cite{SWEBOK2024}). Table~\ref{tab:papers_by_category_ai} shows secondary studies grouped by their AI scope, table~\ref{tab:papers_by_domain_scope} shows secondary studies grouped by SE scope, table~\ref{tab:topics-distribution-clean} shows the distribution of SE KAs, and Figure~\ref{fig:ai_se_intersection} shows intersection of AI and SE scope in secondary studies.

In terms of AI scope, the majority of secondary studies focus on LLMs specifically (21 out of 30, see Table~\ref{tab:papers_by_category_ai}). A smaller number review under broader deep learning (DL) scope (5), general machine learning (ML) scope (2), a combination of ML with non-AI scope (1), or generic AI scope (1). LLMs are reviewed in all of these papers, even under broader scopes where they are treated as one technique among several. The dominance of specifically LLM-oriented reviews (21/30) can indicate that the research community has focused on generative AI, with broader scoped surveys used to capture a broader perspective of research in the area.


In terms of SE scope, most studies focus on \textbf{Software Construction} (25 studies, see Table~\ref{tab:papers_by_domain_scope}). This is to be expected since code generation, for example, and related CGTs most directly contribute to software construction activities. Other frequent scopes include \textbf{Software Maintenance} (14), \textbf{Software Security} (11), and \textbf{Software Testing} (7). Less frequently covered are \textbf{Software Design} (5), \textbf{Software Requirements} (5), and \textbf{Software Engineering Management} (3), which are typically discussed in studies with broad KA scope (in addition to "Software Construction, for example") and may contribute little or nothing to CGTs themselves. Three studies could not be mapped to any SWEBOK KA, and only isolated coverage was found for Software Quality (2), Software Configuration Management (1), and Software Engineering Operations (1). Overall, the SE scope shows a strong concentration on software construction, maintenance, and security with software testing following. In terms of SE scope topic distribution, the majority of secondary studies are narrow in scope: 12 cover only 1 SE KA (see Table~\ref{tab:topics-distribution-clean}), a group of 11 papers span 2–4 topics, and only 4 studies have broad scope (6–8 topics). These broader studies typically include higher-level KAs (e.g., requirements, design, or management) in addition to, rather than instead of, construction and maintenance-oriented topics.

When considering AI and SE scopes together, several patterns emerge:

\begin{itemize}
    \item LLM-focused studies seem to dominate \emph{Software Construction}, with 17 LLM-focused studies addressing construction tasks such as code generation and completion (see Figure~\ref{fig:ai_se_intersection}). Ten of these also intersect with \emph{Software Maintenance} (e.g., program repair) and 9 with \emph{Software Security} (e.g., vulnerability repair).
    \item Likewise, DL-oriented reviews (5 studies) also primarily map to \emph{Software Construction}, with less frequent overlap in \emph{Software Maintenance} and \emph{Software Security}.
    \item ML-scoped reviews (2 studies) are narrower, typically addressing tasks in \emph{Software Construction} and \emph{Software Maintenance}.
    \item Again, a similar trend is observed in "ML \& non-AI" and "AI" studies (2) spanning \emph{Software Construction}, \emph{Software Maintenance}, and \emph{Software Security}.
\end{itemize}
Taken together, these observations suggest that the dominance of construction, maintenance, and security focused SE scope is largely invariant to AI scope.

To sum up, the majority of secondary studies are LLM-focused (21). In terms of SE scope, the major topics covered seem to be "Software Construction" (25), "Software Maintenance" (14), and "Software Security" (11). Half of the studies (15) are narrowly focused (covering 1-2 SE KAs at most) with fewer studies (4) having very broad scope (covering 7-8 SE KAs). The intersection analysis reveals a strong LLM $\times$ Software Construction cluster, indicating that generative models are predominantly investigated in tasks directly related to producing code artifacts. Secondary clusters are visible in LLM $\times$ Software Security and LLM $\times$ Software Maintenance, that could indicate growing interest in applications to vulnerability repair, and long-term code evolution.

\subsubsection{Answering RQ1}
\label{sec:rq1_discussion}
In this section, we synthesize our RQ1a-RQ1d findings and answer the overarching RQ1: \emph{What is the landscape of secondary studies that report (at least in part) on LLM-based CGTs?}

Answering to RQ1, the landscape of secondary studies on LLM-based CGTs can be characterized as follows: 
\begin{itemize}
    \item it has quickly expanded (the number of studies tripled from 2023 to 2024), moving from exploratory studies to more focused systematic studies (0 self-reported SLRs/SMSs in 2022/2023 to 83\% of SLRs/SMSs in 2025);
    \item it shows signs of stabilization and redundancy in the number of primary studies (while the number of secondary studies tripled from 2023 to 2024, the number of primary studies they cover increased from 1491 to 1724 only);
    \item it is maturing in venue quality, with substantial presence in Q1 journals (12/30 studies);
    \item it is dominated by LLMs-focused studies and Software Construction, with secondary KAs in Software Maintenance and Software Security;
    \item it remains narrow in SE scope for most studies (15/30), with only a handful of reviews providing broad, integrative coverage (4/30).
\end{itemize}

\subsection{RQ2: What HELM measures of LLM-based CGTs are reported in secondary studies?}
\begin{table*}[ht]
\centering
\caption{Mapping of secondary studies to HELM measures and Scenarios. A $\checkmark$ indicates presence.}
\label{tab:helm-scenarios}
\scriptsize
\resizebox{0.75\textwidth}{!}{
\newcommand{\rot}[1]{\rotatebox{90}{\strut #1}}

\begin{tabular}{l ccccccc|cccc|ccc}
\toprule
& \multicolumn{7}{c|}{\textbf{HELM measures}} & \multicolumn{4}{c|}{\textbf{Scenario}} & \multicolumn{3}{c}{\textbf{Totals}} \\
\cmidrule(lr){2-8} \cmidrule(lr){9-12} \cmidrule(lr){13-15}
\textbf{Study} &
\rot{Accuracy} & \rot{Calibration} & \rot{Robustness} & \rot{Fairness} & \rot{Bias} & \rot{Toxicity} & \rot{Efficiency} &
\rot{Task} & \rot{Language} & \rot{Application Type} & \rot{Application Domain} &
\rot{HELM} & \rot{Scenario} & \rot{Overall} \\
\midrule
S1  & \checkmark &            & \checkmark &           & \checkmark & \checkmark & \checkmark & \checkmark &            &             &             & 5 & 1 & 6 \\
S2  &            &            & \checkmark &           &            &            &            & \checkmark &            &             &             & 1 & 1 & 2 \\
S3  &            &            &            &           &            &            &            & \checkmark &     \checkmark       &             &             & 0 & 2 & 2 \\
S4  & \checkmark &            &            &           &            &            &            & \checkmark &            &             &             & 1 & 1 & 2 \\
S5  & \checkmark &            & \checkmark &           &            &            & \checkmark & \checkmark & \checkmark &             &             & 3 & 2 & 5 \\
S6  & \checkmark &            &            &           &            &            & \checkmark & \checkmark & \checkmark & \checkmark  &             & 2 & 3 & 5 \\
S7  & \checkmark &            & \checkmark &           &            &            & \checkmark & \checkmark & \checkmark &             &             & 3 & 2 & 5 \\
S8  & \checkmark &            & \checkmark &           &            & \checkmark & \checkmark & \checkmark &            &             &             & 4 & 1 & 5 \\
S9  &            &            & \checkmark &           &            &            & \checkmark & \checkmark &            &             &             & 2 & 1 & 3 \\
S10 & \checkmark &            &            &           &            &            &            & \checkmark & \checkmark &             &             & 1 & 2 & 3 \\
S11 &            &            &            &           &            &            &            & \checkmark &            &             &             & 0 & 1 & 1 \\
S12 & \checkmark &            & \checkmark &           &            &            & \checkmark & \checkmark & \checkmark &             &             & 3 & 2 & 5 \\
S13 & \checkmark &            & \checkmark &           &            & \checkmark &            & \checkmark &            &             &             & 3 & 1 & 4 \\
S14 & \checkmark &            &            &           &            &            & \checkmark & \checkmark &            &             &             & 2 & 1 & 3 \\
S15 &            &            & \checkmark &           &            & \checkmark &            & \checkmark &            &             &             & 2 & 1 & 3 \\
S16 & \checkmark &            &            &           &            &            & \checkmark & \checkmark &            &             &             & 2 & 1 & 3 \\
S17 & \checkmark &            & \checkmark &           &            & \checkmark &            & \checkmark &            &             &             & 3 & 1 & 4 \\
S18 & \checkmark &            & \checkmark &           &            &            &            & \checkmark &            &             &             & 2 & 1 & 3 \\
S19 & \checkmark &            & \checkmark &           &            &            & \checkmark & \checkmark &            &             &             & 3 & 1 & 4 \\
S20 & \checkmark &            & \checkmark &           & \checkmark &            & \checkmark & \checkmark &            &             &             & 4 & 1 & 5 \\
S21 &            &            &            &           &            &            &            & \checkmark &            &             &             & 0 & 1 & 1 \\
S22 &            &            &            &           &            &            &            & \checkmark &            &             &             & 0 & 1 & 1 \\
S23 & \checkmark &            & \checkmark &           &            & \checkmark & \checkmark & \checkmark &            &             &             & 4 & 1 & 5 \\
S24 &            &            & \checkmark &           &            & \checkmark &            & \checkmark & \checkmark &             &             & 2 & 2 & 4 \\
S25 & \checkmark &            & \checkmark &           &            &            & \checkmark & \checkmark &            & \checkmark  &             & 3 & 2 & 5 \\
S26 &            &            &            &           &            &            &            & \checkmark & \checkmark &             &             & 0 & 2 & 2 \\
S27 &            &            &            &           &            & \checkmark &            & \checkmark &            &             &             & 1 & 1 & 2 \\
S28 &            &            &            &           &            &            &            &            &            &             &             & 0 & 0 & 0 \\
S29 &            &            &            &           &            &            & \checkmark & \checkmark &            &             &             & 1 & 1 & 2 \\
S30 & \checkmark &            &            &           &            &            &            & \checkmark &            &             &             & 1 & 1 & 2 \\
\midrule
\textbf{Total} 
& \textbf{18} & \textbf{0} & \textbf{16} & \textbf{0} & \textbf{2} & \textbf{8} & \textbf{14}
& \textbf{29} & \textbf{8} & \textbf{2} & \textbf{0} \\
\bottomrule
\end{tabular}}
\end{table*}

\begin{table}[ht]
\centering
\caption{Summary of evidence for HELM measures, showing number of studies, quality distribution (based on Total QA score), and representative study IDs.}
\label{tab:rq2_qa_measures}
\resizebox{\textwidth}{!}{\begin{tabular}{lclll}
\toprule
\textbf{Measure} & \textbf{\# Studies} & \textbf{High [3 - 4]} & \textbf{Medium [2 - 3)} & \textbf{Low [0 - 2)} \\
\midrule
Accuracy   & 18 & 9 (S1, S8, S10, S13, S18, S19, S20, S23, S25) & 4 (S7, S12, S14, S30) & 5 (S4, S5, S6, S16, S17) \\
Robustness & 16 & 8 (S1, S8, S13, S18, S19, S20, S23, S25)      & 4 (S7, S9, S12, S24)  & 4 (S2, S5, S15, S17) \\
Efficiency & 14 & 7 (S1, S8, S19, S20, S23, S25, S29)           & 4 (S7, S9, S12, S14)  & 3 (S5, S6, S16) \\
Toxicity   &  8 & 5 (S1, S8, S13, S23, S27)                     & 1 (S24)               & 2 (S15, S17) \\
Bias       &  2 & 2 (S1, S20)                                    & 0                     & 0 \\
\bottomrule
\end{tabular}}
\end{table}

To answer RQ2, we have extracted HELM measures" from the secondary studies (see Table~\ref{tbl:sec3-data-rq1}). As can be seen from Table~\ref{tab:helm-scenarios}, no HELM-related data items were extracted from 6 secondary studies. The majority of the remaining 24 studies (20), touched on 1-3 HELM measures, whereas just a few touched on 4-5 measures (4 studies). Across the 24 secondary studies, the most frequently reported HELM measures were Accuracy (18 studies), Robustness (16), and Efficiency (14), see Table~\ref{tab:helm-scenarios}. Toxicity was considered in only 8 studies, Bias in 2, while Calibration and Fairness were not explicitly addressed. In the following sections we synthesize findings for each HELM measure, taking into account papers' quality based on QA scores (see Table~\ref{tab:qa_scores}), where [0 - 2) is low quality, [2 - 3) medium quality, and [3 - 4] high quality as summarized in Table~\ref{tab:rq2_qa_measures}.

\subsubsection{Accuracy}
According to our extracted data, accuracy is the most extensively documented measure of LLM performance in CGTs. It is discussed in 18 studies (see Table~\ref{tab:helm-scenarios}). Of these, 9 are high quality (Total QA 3--4), 4 are medium quality [2 - 3), and 5 are low quality (<2), see Table~\ref{tab:rq2_qa_measures}.

Eight studies (mostly medium-low quality papers) report improvements in accuracy or high accuracy for LLM-based CGTs (S5, S6, S7, S12, S14, S16, S19, S30). These gains are often tied to specific benchmarks and tasks and improvements are in comparison to traditional approaches or older DL models, such as recurrent neural networks (RNN), for example. Several works (S7, S16, S19) show that LLMs achieve high pass@k (a widely used accuracy metric that measures the probability that at least one of the top-$k$ generated solutions for a given task is correct, typically verified against unit tests) scores on HumanEval \cite{chen2021evaluating} and MBPP\cite{austin2021program}, indicating strong functional correctness on short program synthesis tasks. S5 (low quality) report accuracy improvements in program repair benchmarks, where models correctly patch a large fraction of buggy programs. In testing scenarios, S12 and S14 (medium quality) show models generating unit tests that achieve higher coverage than prior baselines.

Seven studies (mostly high quality) (S1, S8, S10, S12, S13, S20, S25) adopt a more cautious stance, questioning whether reported accuracy reflects genuine generalization. Several high-quality contributions (S1, S20) warn that benchmark scores may be inflated by dataset issues. For instance, S1 notes that code duplication and data leakage can lead to overly optimistic accuracy estimates. S20 emphasizes that accuracy metrics often fail to capture semantic correctness, producing syntactically correct but semantically incorrect code. Other high-quality papers (S10, S13) further stress that accuracy varies substantially across tasks and languages, with the same model performing well on HumanEval but poorly on more diverse or real-world datasets. S12 (medium quality) reports modest success in automated program repair, where most generated patches remain incorrect.

Four studies (mostly high quality) explicitly report poor accuracy outcomes (S10, S18, S23, S30). Here, high-medium quality works (S10, S18, S30) demonstrate that models fail on less-represented programming languages and complex integration tasks, with accuracy dropping well below levels required for deployment (S23).

Several recurring themes are reported in the literature that might affect the accuracy of LLM-based CGTs. Three studies (S4, S6, S10) emphasize fine-tuning as an effective way to improve accuracy as compared non-fine-tuned baseline models. Supplying additional context (such as APIs, higher-level project information) (S1, S20) and using few-shot prompting (S6, S10) are also supposed to improve accuracy. The view on the number of model's parameters is mixed: S4 suggests that models with higher number of parameters achieve better effectiveness, whereas S8 highlights that smaller models can outperform larger ones. Finally, the usage of decoder-only architecture (S4) and larger training dataset size (S19) were linked to improved LLM-based CGTs. 

Overall, across the 18 studies, 8 lower-quality studies emphasize improvements, whereas 11 higher-quality studies either adopt a critical cautious perspective (7 studies) or report poor accuracy (4 studies). Several methods such as fine-tuning, providing more context and using few-shot prompting were reported as potentially increasing the accuracy of LLM-based CGTs.

\subsubsection{Robustness}
According to our extracted data, robustness is discussed in 16 studies (S1, S2, S5, S7--S9, S12, S13, S15, S17--S20, S23--S25). Of these, 8 are high quality (Total QA 3--4), 4 are medium quality [2--3), and 4 are low quality (<2), see Table~\ref{tab:rq2_qa_measures}. Across these studies, a consistent view emerges: robustness of LLM-based CGTs is frequently overstated and remains fragile under realistic conditions. 

Eight high-quality studies (S1, S8, S13, S18, S19, S20, S23, S25) provide the strongest and most general evidence for this conclusion. S1 notes that robustness is overstated when training and evaluation datasets overlap, due to duplication / data leakage. S13 reports that widely used benchmarks do not capture real-world scenarios, therefore limiting robustness claims. S20 defines robustness in terms of generalizability and finds that performance does not transfer consistently across tasks, datasets, or domains. S8 describes that robustness must be considered across multiple software engineering activities such as package migration, for example, not only in isolated benchmarks. S23 highlights that both dangers of external attacks on the generated code and internal model limitations can undermine reliable behavior, requiring human oversight. S25 reports that decoding parameters of LLMs such as temperature settings strongly affect output stability.  

Four medium-quality studies (S7, S9, S12, S24) reinforce these findings in more specific contexts. S7 reports that applying off-the-shelf LLMs directly reduces stability in program repair and requires additional heuristics. S9 emphasizes the importance of robustness to adversarial attacks as LLMs are applied more widely in software engineering. S12 shows that test-driven program repair is affected by “test overfitting,” where generated patches pass existing tests but fail on unseen cases, thus limiting generalization. S24 compares robustness across programming languages and notes that results obtained in Python often transfer only partially to C.  

Four low-quality studies (S2, S5, S15, S17) report similar robustness concerns, albeit with narrower scope. S2 finds that improvements in test generation accuracy do not remove robustness issues, as generated tests fail to adapt when requirements change. S15 raises robustness concerns about data poisoning, where malicious triggers in training data compromise downstream behavior. S5 and S17 note that small changes in context or task formulation can reduce performance significantly.  

Overall, the majority of papers here (including all high-quality) are cautious regarding issues with robustness of LLM-based CGTs. High-quality papers emphasize that robustness is often overstated by benchmark-limited evaluations and is sensitive to data quality and evaluation design. Medium and low-quality studies confirm robustness issues in specific scenarios such as program repair and test generation.

\subsubsection{Efficiency}
According to our extracted data, efficiency is discussed in 14 studies (S1, S5--S9, S12, S14, S16, S19, S20, S23, S25, S29). Of these, 7 are high quality (Total QA 3--4: S1, S8, S19, S20, S23, S25, S29), 4 are medium quality [2--3) (S7, S9, S12, S14), and 3 are low quality (<2) (S5, S6, S16), see Table~\ref{tab:rq2_qa_measures}. Overall, these studies report efficiency issues associated with application of LLMs. 

Seven high-quality studies (S1, S8, S19, S20, S23, S25, S29) primarily discuss costs of LLM-based application to CGTs and possible improvements to address efficiency issues. Particularly, 6 studies (S8, S19, S20, S23, S25, S29) mention hardware /computational costs (in terms of time/space required), economical costs (such as high costs of running GPUs), and energy costs (including the impact of CO2 emissions and green energy). Some of the possible solutions discussed involve improving training processes (S1, S19) and utilizing hardware acceleration capabilities (S8, S19).  

Four medium-quality studies (S7, S9, S12, S14) echo these concerns. S12 and S14 highlight computational and economical costs associated with LLM-based CGT application, whereas S9 note limited applicability on resource-constrained hardware such as mobile applications, for example. Interestingly, S7 shows that fine-tuning LLMs is still faster and more efficient than training older DL models (non-pretrained) from scratch.  

Likewise, three low-quality studies (S5, S6, S16) discuss efficiency in similar manner: S5 and S16 highlight all types of costs associated while S6 questions efficiency benefits of LLM-based CGTs, integrated into interactive development environments (IDE).  

Overall, the majority of studies (across all quality groups) highlight efficiency issues (such as computational / hardware / economical / energy / effort costs) associated with application of LLM-based CGTs. Several studies suggest potential improvements in terms of training strategies and hardware acceleration.

\subsubsection{Toxicity}
Toxicity is discussed in 8 studies (S1, S8, S13, S15, S17, S23, S24, S27). Of these, 5 are high quality (Total QA 3--4: S1, S8, S13, S23, S27), 1 is medium quality [2--3) (S24), and 2 are low quality (<2) (S15, S17), see Table~\ref{tab:rq2_qa_measures}.  

Five high-quality studies (S1, S8, S13, S23, S27) report on risks related to harmful, biased, or unsafe outputs produced by LLMs in code-generating contexts. S8 and S23, particularly, mention privacy concerns when private data (such as e-mails, passwords, and usernames for example) can leak from training datasets into code generated by LLMs. S1 highlights that toxicity concerns arise when the quality of identifier names and function names is unaccounted (echoed by S13 for general code vulnerabilities in training datasets) and leak into generated code, increasing vulnerability chances. S27 provides evidence that generated code can be substantially more vulnerable than code created by humans. Two low-quality studies (S15, S17) also raise toxicity concerns in terms of private data leakage into generated code; they also mention copyrighted code fragments appearing in generated data. As a potential solution, one medium-quality study (S24) suggests that vulnerability issues in generated code can be somewhat addressed with few-shot prompting giving secure examples.

Overall, the evidence indicates that toxicity is a less frequently studied measure compared to accuracy or robustness, but it is consistently recognized as a risk. Studies emphasize that toxicity remains unresolved and must be evaluated when LLMs are applied to code-generating tasks, primarily linking these issues to training datasets.

\subsubsection{Bias}
Bias is discussed in just 2 studies (S1, S20), both are high quality (Total QA 3--4), see Table~\ref{tab:rq2_qa_measures}.  

S1 identifies bias resulting from duplicated or cloned code instances in training datasets. The study reports that such data artifacts can bias the model towards highly represented classes, which is described as a form of long-tailed distribution problem. This type of bias can distort model behavior by over-emphasizing common patterns while under-representing rarer but important ones. S20 reports that reliance on limited or biased datasets may cause LLMs to inherit systematic biases, leading to biased or inaccurate predictions in code generation and completion tasks. The study stresses that dataset design and curation are critical to mitigate such issues.  


\subsubsection{Answering RQ2} ("What HELM measures of LLM-based CGTs are reported in secondary studies?") across secondary studies, the \textbf{HELM measures} of LLM-based CGTs are: 
\begin{itemize}
    \item \emph{Accuracy}: Mixed effects reported. While the majority of studies recognizes the effectiveness of LLM-based CGTs on standard benchmarks such as HumanEval and MBPP, the higher-quality studies remain more cautious emphasizing poor/questionable accuracy on a) less-represented languages, b) integration-heavy tasks, and c) real-world scenarios;
    \item \emph{Robustness}: widespread fragility has been reported under dataset/prompt/configuration changes and limited transfer across tasks/datasets/languages;
    \item \emph{Efficiency}: the majority of studies highlight recurring limitations due to computational, economic, and energy costs, with partial mitigation via training and hardware acceleration strategies;
    \item \emph{Toxicity}: this effect is addressed in fewer studies, but consistent recognition of risk are reported regarding privacy of data in generated code and its security (susceptibility to vulnerabilities);
    \item \emph{Bias}: reported in a very limited number of studies, but high-quality evidence linking biased outcomes to data artifacts and distributional imbalance of code (e.g. prevalence of clones). 
\end{itemize}

\subsection{RQ3: What HELM scenarios of LLM-based CGTs are reported in secondary studies?}
To answer RQ3 we extracted data items related to HELM scenario (see Table~\ref{tab:helm-scenarios}). As can be seen only one study (S28) did not provide any Scenario-related data items: the majority of the remaining 29 studies provided 1-2 scenario-related data items (28 studies) and just one study provided 3. Almost all studies (29) reported the task(-s) under consideration, while only a minority specified the languages (8) or application types (2), and none provided details on the application domain. 


Secondary studies report a wide range of code-related tasks, often using heterogeneous or overlapping terminology. Table~\ref{tab:tasks_mapping} presents the original task labels as they appear in the studies, their mapping to our CGT definitions (Section~\ref{sec:def-cgts}), and the corresponding counts of studies in which these tasks are reported.

When tasks are consolidated according to our definitions, \emph{Patch generation / repair} emerges as the most frequently reported CGT (22 studies), aggregating terms such as program repair, vulnerability repair, bug repair, and bug fixing. This is followed by \emph{code generation} (20 studies), and then by \emph{code translation} and \emph{code completion} (9 and 8 studies, respectively). Other CGTs, including \emph{program synthesis} and \emph{test generation} (reported under several synonymous labels), appear less frequently, while \emph{refactoring} is mentioned in only a small number of studies. Some rare task labels (e.g., “code co-evolution,” “code editing,” “method name generation,” and “testing repair”) cannot be mapped directly to our CGT definitions but are closely related to refactoring or repair activities.

\begin{table}[h]
\centering
\caption{Original task terms from secondary studies, mapped to canonical categories; counts are case-insensitive and tasks are expanded from comma-separated lists (sorted by frequency).}
\label{tab:tasks_mapping}
\begin{tabular}{l l r}
\toprule
\textbf{Original term (as reported)} & \textbf{Mapping to definitions} & \textbf{Count} \\
\midrule
code generation & Code generation & 20 \\
code translation & Code translation & 9 \\
code completion & Code completion & 8 \\
program repair & Patch generation / Repair & 8 \\
vulnerability repair & Patch generation / Repair & 6 \\
program synthesis & Program synthesis & 4 \\
test generation & Test generation & 4 \\
code editing & --- (not mapped) & 2 \\
refactoring & Refactoring & 2 \\
test case generation & Test generation & 2 \\
automated bug fixing & Patch generation / Repair & 1 \\
bug fix & Patch generation / Repair & 1 \\
bug repair & Patch generation / Repair & 1 \\
code co-evolution & --- (not mapped) & 1 \\
patch generation & Patch generation / Repair & 1 \\
program translation & Code translation & 1 \\
programming error repair & Patch generation / Repair & 1 \\
security vulnerability repair & Patch generation / Repair & 1 \\
software bug repair & Patch generation / Repair & 1 \\
software vulnerability repair & Patch generation / Repair & 1 \\
testing repair & --- (not mapped) & 1 \\
unit test case generation & Test generation & 1 \\
method name generation & --- (not mapped) & 1 \\
\bottomrule
\end{tabular}
\end{table}

A wide range of programming languages is represented. The most frequently mentioned languages include Java, Python, C, C++, and JavaScript, with several studies reporting multi-language contexts. Beyond these mainstream languages, additional coverage includes C\#, SQL, PHP, Kotlin, Rust, Go, Verilog, Solidity, Ruby, and OCaml. Studies also note domain-specific and low-resource languages, as well as formal languages such as Isabelle/HOL. Also beyond traditional programming languages, Excel and Power Fx are mentioned.

The set of application types is more heterogeneous and reported in only 2 studies. One study mentions low-code applications, highlighting LLM support for end-user programming. Another reports a broad set of software systems, including mobile applications, DL libraries, compilers, SMT solvers, autonomous driving systems, cyber-physical systems, toolchains, JavaScript engines, quantum computing platforms, and video games.

Answering \emph{RQ3: What HELM scenarios of LLM-based CGTs are reported in secondary studies?} we note that secondary studies mostly report tasks, sometimes languages, and rarely application types, but never domains. Having that said, secondary studies report scenarios spanning a wide range of tasks (most prominently code generation, patch generation and repair, code completion, and code translation), applied across multiple programming languages (with strong representation of Java, Python, C, C++, and JavaScript, and additional coverage of domain-specific, low-resource, and end-user languages), and targeting diverse application types (from low-code platforms to compilers, mobile apps, cyber-physical systems, and quantum computing platforms). 

\subsection{RQ4: What challenges in applying or integrating LLM-based CGTs into software engineering workflows are identified in secondary studies?}

\begin{table}[t]
\centering
\caption{Papers grouped by integration/application challenge category (distinct-paper counts). High: QA $\in[3,4]$, Medium: $[2,3)$, Low: $[0,2)$.}
\label{tbl:challenges_results}
\resizebox{\textwidth}{!}{\begin{tabular}{l r l l l}
\toprule
\textbf{Category} & \textbf{Count} & \textbf{High [3-4]} & \textbf{Medium [2-3)} & \textbf{Low [0-2)}\\
\midrule
Economics & 10 & 5 (S13, S19, S23, S25, S29) & 4 (S7, S9, S12, S14) & 1 (S16)\\
Tooling \& Workflow & 7 & 3 (S18, S23, S27) & 2 (S21, S24) & 2 (S6, S17)\\
Evaluation \& Benchmark Validity & 7 & 6 (S1, S8, S10, S13, S20, S29) & 0 & 1 (S5)\\
Data \& Context & 6 & 4 (S8, S19, S20, S25) & 0 & 2 (S15, S17)\\
People \& Process & 5 & 3 (S10, S18, S27) & 2 (S12, S30) & 0\\
Security \& Safety & 5 & 2 (S8, S23) & 2 (S9, S24) & 1 (S15)\\
Model/Training & 4 & 3 (S8, S20, S25) & 1 (S9) & 0\\
Legal \& Privacy & 4 & 2 (S23, S25) & 0 & 2 (S15, S16)\\
\midrule
Not provided & 7 & 2 (S11, S26) & 1 (S22) & 4 (S2, S3, S4, S28)\\
\bottomrule
\end{tabular}}
\end{table}

As can be seen from Table~\ref{tbl:challenges_results} the majority of studies discuss challenges related to Economics (5 high, 4 medium, 1 low), Tooling \& Workflow (3 high, 2, medium, 2 low), and Evaluation \& Benchmark Validity where paper quality is the strongest (6 high, 0 medium , 1 low). These are followed by Data \& Context (4 high, 0 medium, 2 low), People \& Process (3 high, 2 medium, 0 low), and Security \& Safety (2 high, 2 medium, 1 low) categories. Less frequently discussed are challenges in Model/Training (3 high, 1 medium, 0 low) and Legal \& Privacy (2 high, 0 medium, 2 low). Finally, 7 studies reported no challenges, including 2 high-quality papers.

\paragraph{Economics} Across all studies in this category, there seems to be a consistent mention of computational costs and resource constraints/limitations. The former are mentioned in 3 high-quality studies (S13, S23, S29), 3 medium quality (S7, S12, S14), and one low-quality (S16). The computational costs are primarily associated with various stages of LLM training (pre-training, fine-tuning) (S14, S16, S23, S29), inference (S7, S12, S16), and the need to run larger and more computationally expensive benchmarks (S13). Additionally, financial costs are mentioned if off-the-shelf solutions such as GPT-4 are employed (S12). Challenges arising due to resource limitations/constraints are reported in 4 high-quality (S19, S23, S25, S29) and 1 medium quality study (S9). Here, large sizes of LLMs require more computational power, storage, and can affect latency and energy consumption (S19, S25). This is particularly pronounced in resource-limited environments such as on mobile devices where LLMs can struggle to execute, whereas limiting and reducing models' sizes would result in decreased effectiveness (S23). Training can also be limiting for individuals and organizations without the access to resource-rich infrastructure (S23).

\paragraph{Tooling \& Workflow} Across studies in this category, two recurrent issues are mentioned: insufficient integration of LLM-based CGTs into existing tools and pipelines, and the lack of operational controls and responsiveness. The lack of integration with existing tools is highlighted by three studies: 1 high-quality (S18), 1 medium (S24), and 1 low quality (S6). They report the lack of integration with existing IDEs/version control tools/ other existing traditional approaches. One medium-quality study (S21) and two high-quality studies (S23, S27) note the absence of accessible controls such as temperature parameters (an LLM parameter controlling the randomness of its output), for example. High-quality evidence further points to \emph{responsiveness and throughput} as factors for workflow latency reduction, concurrent-user scaling, and memory management to support interactive user workflow (S23). Finally, high-quality studies stress \emph{comprehension and personalization}: suggestions must be reviewable and understandable by a developer, and tools should adapt to individual coding styles and integrate with developer utilities (S23, S27). A low-quality study adds that \emph{adaptive mechanisms} are needed to address context-specific hallucinations (S17).

\paragraph{Evaluation \& Benchmark Validity} Across all evidence in this category, challenges are mostly supported by high-quality sources (6 high, 1 low). Multiple studies argue that common benchmarks are \emph{unrepresentative} of practical development: they emphasize single languages, curated repositories, simplified bugs, and imbalanced or narrow task scopes, which collectively distort difficulty and real-world scenarios (S1, S8, S13). Also, they report poor \emph{generalizability}: models tuned to a specific dataset or task family transfer poorly across domains, languages, or problem types (S29, S13). Several studies identify specification and measurement issues, noting ambiguous task definitions and single-metric reporting: they propose \emph{comprehensive, multi-metric frameworks} that extend beyond functional correctness to include non-functional qualities and practice-level outcomes, with cross-language coverage and adaptable protocols (S8, S10, S13, S20). Finally, one low-quality source notes practical unavailability of executable test fixtures needed by dynamic metrics, further limiting reliable assessment (S5).

\paragraph{Data \& Context} Here, studies emphasize such challenges as \emph{unfamiliar/private} domains and limitations in how much and what kind of context models can use. High-quality studies report difficulties in capturing local–global dependencies for code completion (S19), challenges adapting to repository and system-level problems beyond function scope (S8), and sensitivity to intent ambiguity and project-specific semantics—requiring additional domain knowledge (S20). They also point to data complexity, that can exceed the ability of LLMs to process (S25). Coverage gaps compound these issues: low-resource and domain-specific languages remain underrepresented, constraining applicability (S8). Lower-quality studies align with these observations, noting mishandling of private APIs/repo-wide reasoning and hard limits from input-token windows (S17), as well as domain shift to unseen libraries/environments with impracticality of full fine-tuning, hence the need for prompting/RAG-style adaptation (S15).

\paragraph{People \& Process} High-quality studies in this category highlight communication problem between a developer and an LLM (S18), existing reliance on human oversight (for example in bug fixing) (S10); and the risk by novice developers to overly rely on LLM-based CGTs outputs, underscoring the need for safety training and usage guidance (S27). Medium-quality studies position LLMs primarily as assistants rather than replacements (S30) and note limited empirical understanding of how practitioners actually use these tools in situ (S12).

\paragraph{Security \& Safety} Studies in this category call for rigorous validation in the development pipeline (e.g., formal methods, alignment to coding standards, developer review) (S23, S8). The threats such as backdoors, data memorization, and model extraction/stealing are reported together with the possible mitigations such as obfuscation and client-side protection (S23). Medium-quality studies reinforce the practical risk (e.g. complex web-application tasks are prone to security weaknesses), whereas adversarial attacks are a concern in mission-critical settings (S24, S9). Low-quality evidence reiterates data poisoning as a training-time hazard that can manifest as systematic defects at inference (S15).

\paragraph{Model/Training} High-quality studies argue for continuous learning to track evolving programming practices (S8), for injecting program structure and domain knowledge via code embeddings, syntax/semantic analyses to improve LLMs (S20), and for automated construction of organization-specific datasets from software repositories to reduce the manual cost of fine-tuning while increasing task fit (S25). Medium-quality study adds to this implying the need for further model tuning or specialization (S9).

\paragraph{Legal \& Privacy} Studies here indicate potential threat of leaking private information: code-generating models can reproduce training snippets containing personally identifiable information or secret material, with non-trivial leakage rates, motivating privacy-aware evaluation and mitigations (S23). Organizational responses are also reported: due to confidentiality obligations, many teams avoid commercial endpoints and prefer open-source models fine-tuned on in-house data (S25). Low-quality sources align with this picture, noting that web-crawled corpora may include sensitive artifacts (e.g., credentials) and code under restrictive licences (e.g., GPL), which models can later emit during code generating task (S16, S15). 

\paragraph{Answering RQ4: What challenges in applying or integrating LLM-based CGTs into software engineering workflows are identified in secondary studies?} The most frequently reported challenges concern economics: substantial computational and financial costs across training, inference, and evaluation, together with hard resource constraints that affect latency, energy, and feasibility on constrained devices. Tooling and workflow issues center on incomplete integration in existing tools and workflows (e.g. IDE), limited operational controls (e.g., temperature (LLM randomness parameter)), responsiveness and throughput requirements for interactive use, and the need for comprehensible, personalized suggestions. The evaluation validity concerns are methodologically strongest and highlight unrepresentative benchmarks, poor cross-domain generalization, contamination, ambiguous task specifications, and calls for comprehensive, multi-metric, cross-language assessment frameworks. Data and context challenges include weak grounding in private/unfamiliar domains, token-window limits for repository-scale tasks, and gaps for low-resource or domain-specific languages. People and process concerns emphasize reliance on human oversight, novice over-trust, and poor communication between LLMs and humans. Security and safety issues span vulnerabilities in generated code and model-level threats (adversarial, poisoning, extraction), requiring human validation. Less frequent are model/training needs (continual/task-aligned adaptation, org-specific data pipelines) and legal/privacy risks from memorization and disclosure of protected content. Notably, 7 studies reported no challenges.

\subsection{RQ5: What future directions for LLM-based CGTs are reported in existing secondary studies?}

\begin{table}[htbp]
\centering
\caption{Mapping of papers to future direction categories, grouped by their quality scores.}
\label{tab:rq5_papers_by_category}
\resizebox{\textwidth}{!}{

\begin{tabular}{lrlll}
\toprule
\textbf{Category} & \textbf{Count} & \textbf{High [3--4]} & \textbf{Medium [2,3)} & \textbf{Low [0,2)} \\
\midrule
Model improvement \& training & 16 &
6 (S8, S11, S18, S19, S20, S25) &
7 (S7, S9, S12, S14, S22, S24, S30) &
3 (S5, S15, S28) \\
Benchmark \& evaluation improvement & 11 &
6 (S8, S13, S18, S19, S23, S29) &
3 (S7, S9, S12) &
2 (S15, S17) \\
Ensembling \& hybridization & 7 &
4 (S10, S20, S25, S29) &
2 (S7, S14) &
1 (S17) \\
Advancing experimental research & 7 &
2 (S1, S29) &
4 (S7, S12, S22, S30) &
1 (S5) \\
Security, Privacy \& Trustworthiness & 6 &
3 (S11, S23, S27) &
1 (S22) &
2 (S15, S16) \\
Integration into practice & 5 &
4 (S18, S19, S20, S23) &
0 &
1 (S17) \\
Dataset improvement & 5 &
3 (S1, S20, S29) &
1 (S22) &
1 (S15) \\
\midrule
Not provided & 6 &
1 (S26) &
1 (S21) &
4 (S2, S3, S4, S6) \\

\bottomrule
\end{tabular}
}
\end{table}

The reviewed secondary studies outline several categories of future research directions as shown in Table~\ref{tab:rq5_papers_by_category}. The most frequently mentioned category is \emph{model improvement \& training} (16 studies). Within this category, 6 studies are of high quality (S8, S11, S18, S19, S20, S25), 7 are of medium quality (S7, S9, S12, S14, S22, S24, S30), and 3 are of low quality (S5, S15, S28). Next is \emph{benchmark \& evaluation improvement} (11 studies), with 6 high-quality (S8, S13, S18, S19, S23, S29), 3 medium-quality (S7, S9, S12), and 2 low-quality studies (S15, S17). Further directions include \emph{ensembling \& hybridization} (7 studies: 4 high, 2 medium, 1 low) followed by \emph{advancing experimental research} (7 studies: 2 high, 4 medium, 1 low) and \emph{security, privacy, \& trustworthiness} (6 studies: 3 high, 1 medium, 2 low). The least represented are \emph{dataset improvement} (5 studies: 3 high, 1 medium, 1 low) and \emph{integration into practice} (5 studies: 4 high, 0 medium, 1 low). Finally, 6 studies did not provide explicit future directions (\emph{not provided}: 1 high, 1 medium, 4 low).

\paragraph{Model improvement \& training} A more recurring theme across high-quality papers (S11, S19, S20) is domain-specific integration through specialized datasets, code structures, and programming standards. Others stress interpretability and usability, either via transparent synthesis models (S11) or incorporating readable code and better coding styles into training (S8). Improvement with respect to efficiency is also reported, with calls for computational optimization (S11) and finer-grained code completion (S19). More novel directions include self-repairing models (S20) (where an LLM can analyse and repair/make itself more secure), collaborative open-source development of LLMs (S18), advances in prompt design (S25), and merely increasing the size of LLMs for better effectiveness (S18). Medium-quality studies report improvement directions such as \emph{efficient fine-tuning}: S12 and S9 propose transfer learning, parameter-efficient methods (e.g., prefix-tuning, low-rank adaptation), and targeted fine-tuning. Another direction emphasizes \emph{efficiency at the computational level}, with S14 and S22 suggest improving efficiency through model optimization and leveraging structural code representations such as abstract syntax trees for cross-language generalization. In addition, several studies explore \emph{prompting and context management}, including zero-shot prompt optimization (S24) and advanced preprocessing (S7). Finally, concerns about performance stability and interpretability are raised: S30 notes the lack of rigorous evaluation of many proposed improvements, while S9 suggests incorporating explainable techniques to make predictions more transparent. Low-quality papers (S5, S15, S28) note that \emph{richer code representations} (trees, graphs, data/control-flow features) could better align pre-training with program structure (S5, S15) for model improvement. \emph{Generation scope} such as file-/project-level and the use of chain-of-thought for complex requirements is also highlighted as possible direction (S15). In line with higher quality papers they suggest \emph{domain specialization} (S15) and expanded \emph{prompt engineering}(S28). Finally, they call for \emph{interpretability} in repair models (S5). Overall, domain-specific integration (specialized datasets, code structures, programming standards) seems to be a repeating future research direction towards model improvement across high-medium-low quality studies. Improving prompting, calls for better interpretability, efficiency, and richer code representations are also commonly mentioned. Finally, less commonly mentioned directions here include, the ability of LLMs to self-repair, open-source collaborative development of LLMs, and training with non-functional constraints.

\paragraph{Benchmark \& Evaluation Improvement}
Here across all quality studies, there is broad agreement that current evaluations are narrow and fragmented, motivating a \emph{comprehensive, multi-dimensional framework} to holistically assess LLMs for CGTs (S8, S13, S9, S12, S17, S15). High-quality papers call for frameworks that extend beyond functional correctness to cover code quality, maintainability, performance/efficiency, security, execution stability, and ethics, with cross-language coverage, edge cases, and continuous/living updates (S8, S13). Complementary to this, multiple works advocate \emph{metric standardization} to enable comparability across studies and tasks (S19, S29, S7). Other directions recommend \emph{carefully selected, manually verified test sets} and "live" benchmarks to accurately track progress of LLMs (S18). Finally, studies highlight the need for \emph{scalable evaluation protocols} that balance quantitative metrics with qualitative and human evaluation, to better capture usefulness and robustness in practice (S13).

\paragraph{Ensembling \& Hybridization}
Two trends seem to emerge in this category. First, \emph{model ensembling} (combining multiple LLMs and/or specialized ML models) aims to exploit model heterogeneity to improve performance (S14, S29, S20; high/medium quality). Typical designs include heterogeneous LLM ensembles and LLM+classifier pipelines; expected benefits are variance reduction and specialization across tasks, but they require careful aggregation, cost–benefit analysis, and standardized evaluation to avoid benchmark overfitting. Second, \emph{hybrid techniques} integrate LLMs with traditional SE methods such as static/dynamic analysis, APR heuristics, and domain-specific pipelines (S29, S20, S25, S7, S10, S17; mostly high quality).

\paragraph{Advancing Experimental Research}
In this category, the studies suggest more rigorous and generalizable experimentation. High-quality work suggests expanding study of dataset \emph{data quality}, focusing on the impact towards efficiency, robustness, and security characteristics of generated code (S1). Another study suggest expanding experiments to include industrial-scale evaluations once data access allows (S29). Medium-quality studies emphasize \emph{standardization} parameters, datasets, and experimental settings to enable fair comparisons and to diagnose patch overfitting (S7). Incorporation of \emph{human studies} to assess tool maturity and reliability in practice (S12, S22) is also suggested. Finally, several papers note a coverage bias toward general LLMs and call for targeted evaluations of code-centric models (S30) and for exploratory links across bug classes (S5).

\paragraph{Security, Privacy, and Trustworthiness}
Studies here seem to agree on the need to treat security, privacy, and trustworthiness as important objectives. Security improvement is particularly emphasized with suggestions to treat generated code with caution if deploying (S16, S27). Privacy risks such as data leakage and model memorization are highlighted as potential deployment risks in industry (S15, S22). Beyond correctness, trustworthiness is highlighted as an important property affecting reliability, interpretability of recommendations, and demonstrable generalization across projects and environments (S23). Ethical issues alongside transparency are suggested to be addressed as well (S11). 

\paragraph{Integration into Practice}
High-quality studies emphasize \emph{workflow-centred integration} and \emph{tool-level integration} as the primary routes to integration in practice. Workflow integration entails deployment-ready features and interactive agents, and retrieval-augmented generation (RAG) so that LLM capabilities are integrated within existing development workflows rather than as standalone tools (S18). Tool integration assumes integration into IDEs, code editors, version control, and debugging pipelines (S19, S23). \emph{Context-aware integration} is also mentioned assuming LLM-based CFTs should be more project/task specific (S20, S17).

\paragraph{Dataset Improvement}
In this category studies propose several future directions. S1 (high quality) suggests cleaned versions of datasets to be used for training, removing data smells. S29 and S20 (high quality) suggest \emph{domain-specific datasets} to encode programming-domain knowledge in specialized contexts. Medium-quality work emphasizes \emph{representativeness}: S22 recommends more diverse, real-world datasets to better reflect industrial conditions. S15 (low-quality) proposes manual/automatic cleaning to mitigate data poisoning and suggests \emph{adversarial augmentation} (perturbed examples) to improve robustness. 

\paragraph{Answering RQ5: What future directions for LLM-based CGTs are reported in existing secondary studies?} Secondary studies most consistently recommend the following areas for future research directions (sorted from most to least mentioned):
\begin{itemize}
    \item Model improvement \& training, particularly focusing on domain-aware model improvement;
    \item Benchmark \& evaluation improvement with the help of comprehensive/holistic benchmarks and standardized evaluation; 
    \item Ensembling \& hybridization with either ensembles of ML models (LLMs and other) or hybridization with traditional approaches; 
    \item Advancing experimental research including more rigorous experimental methodology and incorporation of human/industrial studies;
    \item Improving security privacy \& trustworthiness of LLM-based CGTs;
    \item Integration into practice by merging into existing workflows and tools;
    \item Dataset improvement via higher-quality data and incorporation of domain-specific knowledge.
\end{itemize}

\section{Discussion}
\label{sec:discussion}
\subsection{RQ1: Landscape of secondary studies on LLM-based CGTs}
The analysis of secondary studies shows their expansion from 2022 to 2024 followed by sustained, if leveling-off, activity in the first half of 2025. There is a marked shift from exploratory studies (surveys/reviews) toward more systematically-performed studies (SLRs particularly): this could suggest a field transitioning from exploratory mapping of studies to more focused synthesis of studies in the area. In evidence‐based software engineering, such a shift could be interpreted as a maturation signal: once the topic space has been mapped and terminology stabilized, researchers increasingly look into aggregative inference and interpretation rather than mere mapping \cite{kitchenham2007guidelines, petersen2015guidelines}. There also seem to be growing expectations by high-ranked venues for more rigorous synthesis: the concentration of publications in Q1 journals (e.g., TOSEM, TSE, CSI) further reinforces this interpretation.

At the same time, the total number of primary studies reviewed between 2023 and 2024 has increased only by a small margin (despite a tripling of secondary studies). Two non-exclusive explanations can be plausible here: the speed of primary studies appearing has slowed down somewhat and/or cross-study redundancy arising from similar search strings, sources, and inclusion criteria. From a research-efficiency standpoint, this raises risks of duplicated effort without proportional gain. Reusable transparent datasets (with supplementary protocols) could be used to address this \cite{kitchenham2007guidelines, petersen2015guidelines}. Practically, this trend implies that future contributions will likely need to differentiate less by volume of included studies and more by scope refinement, methodological innovation (e.g., quantitative meta-analytic models, bias control \cite{kitchenham2022segress}), or theory-building. It should be noted, however, that the trends above are observed over a relatively short time span (approximately 3.5 years) and should therefore be interpreted as early indicators rather than long-term trajectories.

The dominance of LLM-scoped studies, together with the strong prevalence of SWEBOK’s Software Construction (followed by Maintenance and Security), highlights the application profile of CGTs from an AI perspective: the majority of CGTs are used to construct code and to repair bugs or fix vulnerabilities using LLMs.

Finally, venue patterns provide an interesting quality signal. Journal-published studies show the highest average QA scores, consistent with rigorous peer-review process. However, arXiv-hosted studies display only slightly lower QA on average, indicating that preprint dissemination in this fast-moving domain does not necessarily trade rigor for speed (e.g. during the creation of this publication, one paper (S23) previously hosted on arxiv was accepted in TOSEM). Taken together, the landscape depicts a field that is quickly evolving, maturing methodologically and qualitatively, yet still to consolidate around reusable/transparent protocols and datasets, and broader coverage of less represented SWEBOK KAs that require code generating activities (e.g. Software Testing).

\subsection{RQ2: HELM measures of LLM-based CGTs}
The synthesis of HELM measures reveals several important patterns. First, \emph{measure representation is uneven}. Accuracy is the most frequently reported outcome, while robustness and efficiency are reported less often. Toxicity and bias remain rarely reported, whereas calibration and fairness are not addressed at all (it should be mentioned that the latter two are less applicable to SE context, though).

Although there seems to be a common acknowledgment of accuracy improvements in LLM-based CGTs, higher-quality studies consistently remain cautious regarding its reliability (especially when applied to real-world scenarios) due to dataset contamination and benchmark limitations. This aligns with broader critiques in the literature that reported improvements in benchmark-based performance metrics (e.g., pass@k or success rates on curated code-generating benchmarks) may be inflated by data leakage or narrow evaluation setups \cite{chen2021evaluating,austin2021program}.

Robustness is widely reported as fragile, with performance varying substantially across tasks, languages, and configurations. These findings echo concerns in program repair and software testing studies, for example, where LLM-generated patches often overfit test suites and fail on unseen cases \cite{parvez2021retrieval}. 

Efficiency concerns are systematically reported: computational, economic, and energy costs are highlighted across quality levels, consistent with sustainability concerns in broader natural language processing (NLP) research \cite{strubell2020energy}. 

Toxicity and bias are underexplored but consistently acknowledged as risks, reinforcing conclusions from the AI safety literature that harmful outputs and data biases are persistent challenges \cite{weidinger2022taxonomy}. Calibration and fairness are absent, indicating that evaluation practices in SE are not entirely consistent with those in broader NLP, where these measures are more standard. 

Collectively, the HELM measures suggest that while LLMs achieve measurable accuracy gains on short-form synthesis tasks, these can be fragile and should be cautiously assessed when applying to practice. Efficiency remains a concern, and safety-related risks are insufficiently addressed. Moving towards a more comprehensive benchmark (incorporating several measures) and expanding research into less studied areas could partially address these concerns.

\subsection{RQ3: HELM scenarios of LLM-based CGTs reported in secondary studies}
The scenario evidence reported by secondary studies is mostly reported in terms of the task dimension, with infrequent reporting of language and very rare reporting of application type. Application domain is not reported at all. This can limit the interpretability of LLM-based CGT studies and aggregation synthesis of results and could also undermine the external validity of such studies. The ‘task’ attribute of the scenario seems to have a large number of synonyms (e.g., overlapping repair labels) and, despite our consolidation of commonly reported CGTs based on prior task lists and the LLM-based literature, would benefit from a more widely adopted controlled vocabulary and task normalization to reduce ambiguity and enable meta-analysis and synthesis across studies. At the same time, the broad language coverage reported, spanning general, domain-specific, low-resource, and even end-user languages, suggests that LLM-based CGTs are being evaluated across diverse programming ecosystems. In the future, adopting standardized taxonomies when reporting could improve reproducibility and strengthen claims about generalizability.

\subsection{RQ4: Challenges in applying or integrating LLM-based CGTs}
The economics related challenges seem to prevail: training, adaptation, and inference costs, evaluation at scale, and the latency/energy constraints of deployment environments jointly limit feasibility and scope. Compounding this, the presence of high-quality evidence around evaluation and benchmark validity indicates that current assessment practices insufficiently represent the heterogeneity of real projects (languages, repositories, dependency graphs, non-functional requirements), increasing risks of overfitting to narrow tasks and cherry-picked data.

Tooling \& Workflow challenges suggest insufficient integration with existing systems, limited operational controls, and responsiveness/throughput bottlenecks that impede continuous, interactive use. These observations are in line with prior reports of integration and maintenance concerns in ML systems, where the model is the 'easy' part and the surrounding infrastructure is what determines product viability \cite{sculley2015hidden}. Highly aligned with this concern, People \& Process suggest that the effective use of LLM-based CGTs assumes communicative alignment between models and developers, appropriate safety training for novice developers, and explicit human controls. Data \& Context challenges further highlight that the empirical improvements observed on public benchmarks (e.g. MBPP) may not transfer to real-world codebases without targeted adaptation (prompting, retrieval, or fine-tuning) and expanded context handling.

Finally, Security \& Safety challenges including vulnerabilities of sorts in generated code, adversarial attacks, and model-level risks (such as memorization, extraction) imply that CGTs must be embedded in guarded pipelines (e.g. organizational policy, static/dynamic analysis, code review) and possibly assessed with threat-aware evaluation protocols \cite{carlini2021extracting}. Legal \& privacy risks result in emission of licensed or sensitive content and can imply organizations use in-house/fine-tuned models. Overall, a more successful and widespread integration of LLM-based CGTs into real-world software systems/workflows might require addressing economic concerns, comprehensive multi-metric and multi-language evaluation, infrastructure that enables controllable, explainable, and low-latency interaction, and addressing security, privacy, and people \& process concerns.

\subsection{RQ5: Future directions for LLM-based CGTs}
Model improvement and training is the most commonly reported future direction among the secondary studies. It can suggest that there is still a decent room for improvement of LLM-based CGTs. Particularly, domain-aware adaptation of the models is repeatedly proposed, reflecting a move towards specialized models tuned to organizational codebases and standards. Efficiency, interpretability, and controllability are mentioned as much as accuracy is. This is consistent with prior studies highlighting ML systems value-increase when models are engineered to align with context, data, and downstream tasks rather than optimized in isolation \cite{sculley2015hidden}.

Second mostly reported are recommendations for benchmark and evaluation improvement (multi-metric, cross-language, and real-world scenario based). This can suggest that current benchmarks insufficiently reflect real software work, aligning with similar concerns in broader LLMs for NLP (where, for example, accuracy alone is not sufficient) \cite{liang2022holistic}. The emphasis on ensembling and hybridization (across LLMs and traditional approaches) and on integration into practice can indicate a pivot from standalone models to systems of models and /or other approaches. This is consistent with current existing directions in LLM for SE research, where ensembles, for example, are employed to improve models performance \cite{ahmed2023using}.

Finally, directions around security, privacy, and trustworthiness and dataset improvement, though least-reported, are as important. They can suggest rising concerns regarding security vulnerabilities in existing models, privacy violations, and the impact of poorly constructed datasets towards these issues. Overall, there seems to be calls towards domain-specialized modeling with real-world scenario, multi-dimensional evaluation. Additionally, hybrid/ensemble systems seem to be a viable path towards improving models' performance, whereas security and privacy concerns of LLMs are equally need to be addressed. Notably, comparatively few studies emphasize integration into practice, despite the increasing deployment of LLM-based CGTs in real-world settings. This limited attention suggests that human-in-the-loop concerns remain under-explored relative to model and benchmark-centric directions.

\subsection{Application of LLM toward literature review}
In this work, AI (GPT4o LLM) was used in backward and forward snowballing to screen large corpora of potentially relevant abstracts and the screening results were subsequently cross-checked by the first author (MC). The results indicate that the AI system consistently captured relevant studies with high precision and perfect observed recall in the validation sample. The agreement rate (0.83) was comparable and even slightly exceeding dual human screening during SLRs \cite{perez2020systematic}. Our strategy and results seem to align with existing literature such as a screening strategies proposed by Waffenschmidt et al. \cite{waffenschmidt2019single}, in which one reviewer or model excludes papers while all potential inclusions are retained. Recently, within SE domain, Huotala et al. \cite{huotala2024promise} have shown that GPT-based models can approach human-level performance in screening for systematic reviews, validating the model's use in SE-specific contexts. This approach is further supported by evidence from other recent research in other domains: for example, Nykvist et al. \cite{nykvist2025testing} demonstrated that GPT-4 could achieve perfect recall in screening abstracts for environmental systematic reviews, and Guo et al. \cite{guo2024automated} found similar performance in the medical domain.

\section{Threats to validity}
\label{sec:threats}
In line with Kitchenham’s guidance for systematic reviews in software engineering and SEGRESS reporting \cite{kitchenham2007guidelines,kitchenham2022segress}, we report on threats to construct, internal, and external validity.

\paragraph{Construct validity}
Here threats may arise from how data was used and coded. For example, study type labels (e.g., “SLR”, “SMS”, “survey”) are taken as reported by the authors, which may not always align with formal methodological definitions \cite{kitchenham2007guidelines}. Second, and similarly, multi‐valued non-trivial constructs (e.g., SE scope via SWEBOK v4 KAs \cite{SWEBOK2024}, scenario attributes, and challenge categories) require normalization of heterogeneous terminology. We mitigated this by establishing extraction and coding rules, cross-validating between the authors, and mapping to standard existing definitions/frameworks (e.g. SWEBOK, HELM framework), where possible.

\paragraph{Internal validity}
Selection bias and non-inclusion can occur during database searches, screening, and snowballing. We searched multiple bibliographic sources recommended for SLRs \cite{gusenbauer2020academic} and used a PRISMA-like flow \cite{moher2015preferred} with deduplication, yet different index coverage (i.e, variation in which and how venues, publication types, and metadata are indexed across databases) in paper databases and metadata may still omit eligible studies. For Google Scholar (GS), we used fixed blocks of 50 in relevance order and applied a stop rule (two consecutive blocks with $<5\%$ inclusion; a hard cap of 300 results) informed by best-practice guidance on GS precision and practicality \cite{gusenbauer2020academic, haddaway2015role, bates2017will, walters2009google}. This balances feasibility and sensitivity but may miss low-ranked relevant items. During snowballing we adopted Wohlin’s iterative procedure with a stopping rule ($<5\%$ yield) \cite{wohlin2014guidelines}: setting such a threshold risks early termination and omission of relevant studies. To limit single-reviewer bias in title/abstract screening, we audited a balanced random sample of 30 items with multiple raters and obtained almost-perfect agreement (Fleiss’ $\kappa=0.86$) \cite{fleiss1971measuring,landis1977measurement}. For semi-automated screening of 2,769 snowballed records, we validated GPT-4o decisions on a 100-item random sample, observing precision $=0.77$ (Wilson 95\% CI 0.60–0.89) and recall $=1.00$ (Wilson 95\% CI 0.86–1.00), and Cohen’s $\kappa=0.83$ against human judgment\cite{wallace2010semi,o2015using,waffenschmidt2019single,perez2020systematic}. Despite this, the risks still include GS ranking of artifacts, prompt sensitivity in AI screening, and missed citations due to incomplete reference metadata.

\paragraph{External validity (generalizability)}
Our scope targets secondary studies on LLM-based CGTs published since 2017, written in English, and includes arXiv as the sole source of grey literature. As a result, the findings may not generalize to non-English reviews or to grey literature beyond arXiv.

More broadly, the results characterize patterns and emphases in the surveyed secondary literature, rather than the full body of primary CGT research. Additionally, because several trends are observed over a relatively short time span (approximately 3.5 years), their generalizability beyond the studied time-frame remains uncertain.

\section{Conclusions}
\label{sec:conclusions}
To the best of our knowledge, this tertiary study provides the first systematic synthesis of secondary evidence on Large Language Models (LLMs) applied to code-generating tasks (CGTs) in software engineering. To accomplish this, we leveraged Kitchenham/SEGRESS protocol suggested by Kitchenham et al. for conducting/reporting secondary/tertiary studies in software engineering. Particularly, we systematically searched 5 article databases and Google Scholar, subsequently extending this with snowballing to obtain the final set of 30 relevant secondary studies. To address bias risk in extracting/synthesizing/reporting data we systematically employed extractor+checker and multi-rater strategies, achieving almost-perfect agreements.

The findings portray a rapidly evolving yet unevenly structured field: in less than three years, even the number of secondary studies has increased substantially (for example, going from 6 in 2023 to 17 in 2024), while also marking a methodological shift from exploratory surveys toward systematic literature reviews and mappings (83\% of the latter in 2025). This can suggest maturation of this research area, yet also reveals emerging redundancy and limited coverage of underrepresented software engineering knowledge areas such as testing and management (the majority of studies reviewed seem to focus on LLM-based software construction, n=17).

Across HELM measures, LLM-based CGTs demonstrate notable but uneven progress. Accuracy is mostly discussed (n=18) and gains are well documented on standardized benchmarks but questioned by higher-quality studies that highlight questionable applicability to real-world tasks. Robustness (n=16) remains inconsistent across tasks/datasets/languages, and efficiency (n=14) constraints (computational, economic, and energetic) persist across studies. Toxicity and bias, though under-reported (n=8 and n=2, respectively), consistently appear as systemic risks, primarily linked to training data and uncurated code corpora. The scenarios to which these effects apply mostly involve such tasks as code generation, various code repair related generation, code completion and translation.


The challenges synthesis underscores the barriers to reliable integration of LLM-based CGTs into development workflows. At a higher level, three dominant challenge clusters emerge: economic feasibility (n=10), evaluation validity (n=7), and socio-technical integration issues combining tooling/workflow and people/process concerns (n=12). In particular, high training and inference costs, energy consumption, and hardware and software constraints represent major obstacles to adoption. High-quality studies further emphasize that current benchmarks are often unrepresentative, task definitions ambiguous, and that multi-metric frameworks are required to holistically capture performance, security, and human-in-the-loop factors. The prominence of socio-technical integration challenges—spanning limited IDE integration, poor operational control and responsiveness, trust, and organizational fit—highlights a persistent gap between experimental LLM-based CGTs and deployable software engineering tools. Meanwhile, legal, privacy, and security risks remain insufficiently reported and mitigated.

Future research directions converge toward several priorities. The most widely endorsed involve domain-aware model adaptation through specialized datasets and fine-tuning, multi-dimensional and standardized evaluation frameworks, hybrid and ensemble approaches combining LLMs with traditional SE methods, and the inclusion of human-centered and industrial studies to enhance validity.

\begin{acks}
This work was supported by Science Foundation Ireland grant 13/RC/2094\_2.
\end{acks}

\bibliographystyle{ACM-Reference-Format}
\bibliography{sample-base}


\end{document}